\documentclass[11pt,fleqn]{article}
\usepackage{
feynmp
,rangecite
,graphicx
}
\usepackage{figs}
\newcommand\myappendix{\par
  \setcounter{section}{0}%
  \setcounter{subsection}{0}%
  \gdef\thesection{Appendix \Alph{section}}}
\newcommand{\Htb}[1][]{\ensuremath{H^+\rightarrow t\,\bar{b}}}
\newcommand{\tbH}[1][]{\ensuremath{t\rightarrow b\,H^+}}

\newcommand{\tb}[1][]{\ensuremath{\tan^{#1}\!\beta}}
\newcommand{\ctb}[1][]{\ensuremath{\cot^{#1}\!\beta}}

\newcommand{\mt}[1][]{\ensuremath{m_t^{#1}}}
\newcommand{\mb}[1][]{\ensuremath{m_b^{#1}}}
\newcommand{\Dmb}[1][]{\ensuremath{\Delta\mb[#1]}}

\newcommand{\omb}[1][]{\ensuremath{\overline{m}_b^{#1}}}

\newcommand{\rt}[1][]{\ensuremath{r_t^{#1}}}
\newcommand{\rb}[1][]{\ensuremath{r_b^{#1}}}
\newcommand{\qb}[1][]{\ensuremath{q_b^{#1}}}
\newcommand{\qh}[1][]{\ensuremath{q_{H^+}^{#1}}}

\newcommand{\mw}[1][]{\ensuremath{M_W^{#1}}}
\newcommand{\mH}[1][]{\ensuremath{M_{H^+}^{#1}}}

\newcommand{\mg}[1][]{\ensuremath{M_{\tilde{g}}^{#1}}}
\newcommand{\msb}[2][]{\ensuremath{m_{\tilde{b}_{{#2}}}^{{#1}}}}
\newcommand{\mst}[2][]{\ensuremath{m_{\tilde{t}_{{#2}}}^{{#1}}}}
\newcommand{\MSy}[1][]{\ensuremath{M_{SUSY}^{#1}}}
\newcommand{\sst}[1][]{\ensuremath{\sin^{#1}\!\theta_{\tilde{t}}}}
\newcommand{\cst}[1][]{\ensuremath{\cos^{#1}\!\theta_{\tilde{t}}}}
\newcommand{\ssb}[1][]{\ensuremath{\sin^{#1}\!\theta_{\tilde{b}}}}
\newcommand{\csb}[1][]{\ensuremath{\cos^{#1}\!\theta_{\tilde{b}}}}

\newcommand{\aS}[1][]{\ensuremath{\alpha_s^{#1}}}
\newcommand{\oaS}[1][]{\ensuremath{\alpha_s^{#1}}}

\newcommand{\MS}[1][]{\ensuremath{\overline{\mathrm{MS}}}}

\newcommand{\gsim}{\stackrel{\scriptstyle >}{{ }_{\sim}}}

\newcommand{\dqcd}[1][]{\ensuremath{\delta\Gamma_{QCD}^{#1}}}
\newcommand{\dsusy}[1][]{\ensuremath{\delta\Gamma_{SUSY}^{#1}}}
\newcommand{\dmssm}[1][]{\ensuremath{\delta\Gamma_{MSSM}^{#1}}}

\newcommand{\eq}[1]{(\ref{#1})}
\newcommand{\ov}{\overline}
\newcommand{\lt}{\left}
\newcommand{\nn}{\nonumber \\}
\newcommand{\no}{\nonumber }
\newcommand{\imag}{\mbox{Im\hspace{1mm}}}
\newcommand{\bra}[1]{\langle \, #1 \, | }
\newcommand{\ket}[1]{| \, #1 \, \rangle }
\begin{document}

%%\leftdraft{\today} 

%\phantom{.}
\begin{flushright}
\parbox[t]{2in}{
CERN-TH/2000-009 \\
FERMILAB-Pub-99/367-T \\
hep-ph/9912516
}
\end{flushright}
\vspace{3em}
%\vspace{7em}

\begin{center}
{\Large
{\bf Effective lagrangian for the
     $\mathbf{\bar{t}bH^+}$ interaction in the MSSM
     and charged Higgs phenomenology}
}
\vspace{2em}

\def\thefootnote{\fnsymbol{footnote}}

Marcela Carena,${}^{a}$\footnote{On leave from the Theoretical Physics
  Department, Fermilab, Batavia, IL 60510-0500, USA.}%
\, David Garcia,${}^a$%
\, Ulrich Nierste,${}^{b}$%
\, Carlos E.M. Wagner\phantom{,}${}^{a}$\footnote{On leave from the High
  Energy Physics Division, Argonne National Laboratory, Argonne,
  IL 60439, USA\\
  and the Enrico Fermi Institute, Univ.\ of Chicago, 5640 Ellis, Chicago, IL
  60637, USA.}%
\\[1em]
${}^a$ Theory Division, CERN, CH-1211 Geneva 23, Switzerland\\
${}^b$ Fermi National Accelerator Laboratory, Batavia, IL 60510, USA

\vspace{2em}
\end{center}

\def\thefootnote{\arabic{footnote}}
\setcounter{footnote}{0}

\begin{abstract}
  In the framework of a 2HDM effective lagrangian for the MSSM, we
  analyse important phenomenological aspects associated with quantum
  soft SUSY-breaking effects that modify the relation between the
  bottom mass and the bottom Yukawa coupling. We derive a resummation
  of the dominant supersymmetric corrections for large values of \tb\ 
  to all orders in perturbation theory. With the help of the operator
  product expansion we also perform the resummation of the leading and
  next-to-leading logarithms of the standard QCD corrections. We use
  these resummation procedures to compute the radiative corrections to
  the \tbH, \Htb\ decay rates. In the large \tb\ regime, we derive
  simple formulae embodying all the dominant contributions to these
  decay rates and we compute the corresponding branching ratios. We
  show, as an example, the effect of these new results on determining
  the region of the \mH--\tb\ plane excluded by the Tevatron searches
  for a supersymmetric charged Higgs boson in top quark decays, as a
  function of the MSSM parameter space.
\end{abstract}

\begin{flushleft}

\end{flushleft}
\vspace{2em}

%\begin{flushleft}
%\parbox[t]{2in}{
%CERN-TH/2000-009 \\
%FERMILAB-Pub-99/367-T \\
%hep-ph/9912516
%%%%January 2000
%}
%\end{flushleft}

\newpage
\section{Introduction}

In minimal supersymmetric extensions of the Standard Model (SM), soft
Supersymmetry (SUSY) breaking terms \cite{SSB} are introduced to break SUSY
without spoiling the cancellation of quadratic divergences in the process of
renormalization. These terms must have dimensionful couplings, whose values
determine the scale \MSy, lower than a few TeV, above which SUSY is
restored; they are also responsible for the mass splittings inside the
supersymmetric multiplets. Little is known for sure about the origin of these
SUSY-breaking terms. Upcoming accelerators will test the energy range where we
hope that the first supersymmetric particles will be found. From their masses
and couplings we could learn about the pattern of SUSY-breaking at low
energies, which translates, through the renormalization group equations, into
the pattern of breaking at the scale at which SUSY-breaking is transmitted to
the observable sector. Meanwhile, one can obtain some information on the soft
terms by looking at any low-energy observables sensitive to their values, and
in particular to the Yukawa sector of the theory.

In this work we consider the simplest supersymmetric version of the SM, the
Minimal Supersymmetric Standard Model (MSSM) \cite{Reports}.  We analyse the
limit of a large ratio $v_2/v_1=\tb$ of the vacuum expectation values $v_1$,
$v_2$ of the Higgs doublets.  We show that in this limit a large class of
physical observables involving the Yukawa coupling of the physical charged
Higgs boson can be described in terms of a two-Higgs-doublets model (2HDM)
\cite{hhg} effective lagrangian, with specific constraints from the underlying
MSSM dynamics.

The finding of a charged Higgs boson would be instant evidence for physics
beyond the SM. It would also be consistent with low-energy SUSY, as
all supersymmetric extensions of the SM contain at least a charged Higgs
boson, $H^\pm$.  Current experiments, looking at the kinematical region
$\mH<\mt - \mb$, have been able to place an absolute bound of $\mH>71.0$~GeV
at the 95\% confidence level \cite{LEP} and/or to exclude regions of the
\mH--${\cal BR}(\tbH)$ plane \cite{Dhiman,TEV}.\footnote{See also the study in
  ref.~\protect{\cite{Bo98}}, where it is shown how these bounds are affected
  by some usually overlooked decay modes in the intermediate $\tb\gsim 1$
  region.}  If the charged Higgs mass happens to be greater than the top mass,
future $e^+e^-$, $p\bar{p}$ and even $e^-p$ accelerators will have a chance to
find it \cite{firstHp,Ku92,moreHp}.

Present bounds from LEP on a SM light Higgs boson, $M_{h^{SM}}>105.6$~GeV
\cite{AlOp}, are beginning to put strong constraints on values of \tb\ lower
than a few, a region that can only be consistent with low-energy SUSY
if the third-generation squark masses are large, of the order of a TeV and, in
addition, if the mixing parameters in the stop sector are of the order of, or
larger than, the stop masses. Therefore, the LEP limits give a strong
motivation for the study of the large \tb\ region. The region of large values
of \tb\ is also theoretically appealing, since it is consistent with the
approximate unification of the top and bottom Yukawa couplings at high
energies, as happens in minimal SO(10) models \cite{SO10,copw}. The aim of
this work is to compare, for large values of the \tb\ parameter, the effective
potential results truncated at one loop with the diagrammatic one-loop
computation for the supersymmetric QCD (SUSY-QCD) and electroweak (SUSY-EW)
corrections in the coupling of $\bar{t}bH^+$ \cite{SQCDHtb,Co98,Gu95,Co96}. We
then use the effective potential approach to include a resummation of the
SUSY-QCD and SUSY-EW effects and we show how relevant these higher-order
effects are to the final evaluation of the \Htb\ and \tbH\ partial decay
rates.

\begin{figure}[t]
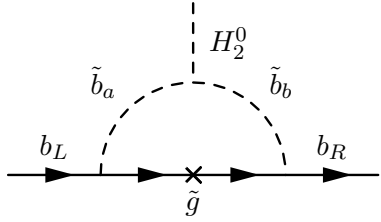

  \begin{center}
    \figvsqcd
  \end{center}
  \caption{\label{fig:vSQCD} One-loop SUSY-QCD diagram contributing to 
    the effective coupling $\Delta h_b$. The solid lines inside the loop
    denote the gluino propagator, the dashed lines correspond to sbottom
    propagators. The cross represents the \mg\ insertion coming from the
    gluino propagator.}
\end{figure}

Although diagrammatic computations of the ${\cal O}(g\aS)$ quantum corrections
to these observables have existed in the literature for several years, either
in the context of a generic two-Higgs-doublets model
\cite{QCDHtb,Dj95,QCDtbH,Cz93},\footnote{For the QCD corrections to the
  neutral Higgs decay rate the reader is referred to \protect{\cite{bl,Ha97}}
  and references therein.} or in supersymmetric extensions of the Standard
Model \cite{SQCDHtb,Co98,Gu95,Co96}, our analysis goes beyond these studies in
the following:
\begin{itemize}
\item It resums leading and next-to-leading logarithms of the type 
   $\aS \log(\mb/\mt)$ or\\ $\aS \log(\mb/\mH)$, because these terms are
   of the same size as the tree-level result.
 \item It includes the potentially large supersymmetric corrections
   responsible for the leading behaviour at large $\tan\beta\geq 10$ values,
   {\it with an improved treatment of the higher-order contributions}
   incorporated into the effective lagrangian: the corrections of order
   $(\aS\mu\tb/\MSy)^n$ are included to all orders $n=1,2,\ldots$ These
   corrections do not vanish if the parameter $\mu$ and the soft SUSY-breaking
   masses are pushed to large values, which is a reflection of the lack of
   supersymmetry in the low-energy theory.
\item It is well suited for numerical evaluation, because it includes all the
  relevant terms, by means of very simple formulae. Therefore, the bulk of the
  quantum corrections can be implemented in a fast Monte Carlo generator.
\end{itemize}
We would like to stress the second point: even for a heavy supersymmetric
spectrum, depending on the ratios and relative signs of the Higgs mass
parameter, $\mu$, and of the soft SUSY-breaking parameters involved, the
supersymmetric QCD and EW corrections can be very large, a situation in which
the higher-order effects are sizeable.

The text is organized as follows. In section~\ref{sec:efflag} we derive the
coefficients of the 2HDM effective lagrangian which are affected by large SUSY
threshold effects. Section~\ref{sec:quantum} provides simple analytical
expressions for the QCD and electroweak quantum corrections to the \tbH\ and
\Htb\ partial decay rates, including the resummation of the large leading and
next-to-leading QCD logarithms and of the potentially large \tb-enhanced SUSY
corrections. Section~\ref{sec:Widths} is devoted to the numerical analysis of
the partial widths, comparing them to the previously existing one-loop results
\cite{Co98,Co96}. To exemplify the importance of these novel computations, we
show in section~\ref{sec:BRs} their effects on the ${\cal BR}$'s of \tbH\ and
\Htb. As an example we study the effects of these results on the limits on the
\mH\ mass derived by the D0 collaboration (similar limits have been obtained
by the CDF collaboration) via the indirect search of the charged Higgs in
$t\bar{t}$ decays. We reserve section~\ref{sec:concl} for our summary and
conclusions.

\section{Effective lagrangian}
\label{sec:efflag}

\subsection{Supersymmetric corrections}
\label{sec:uli:susy}

The effective 2HDM lagrangian contains the following couplings of the bottom
quark to the CP-even neutral Higgs bosons \cite{Ca98}:
\begin{equation}
  \label{eq:efflag}
h_b H_1^0 b\bar{b}+\Delta h_b H_2^0 b\bar{b}\,.
\end{equation}
The $H_2^0b\bar{b}$ tree-level coupling is forbidden in the MSSM. Yet
a non-vanishing $\Delta h_b$ is dynamically generated at the one-loop
level by the diagram of fig.~\ref{fig:vSQCD}.\footnote{There are similar
diagrams involving supersymmetric electroweak quantum corrections, see
section~\protect{\ref{sec:SUSY}}.}

Although $\Delta h_b$ is loop-suppressed, once the Higgs fields $H_{1,2}^0$
acquire their vacuum expectation values $v_{1,2}$, the small $\Delta h_b$
shift induces a potentially large modification of the tree-level relation
between the bottom mass and its Yukawa coupling, because it is enhanced by
$\tb=v_2/v_1$:
\begin{equation}
  \label{eq:massYukawa}
  \mb=h_b v_1 \longrightarrow \mb=v_1\left(h_b+\Delta h_b\tb\right)
                                 =h_b v_1\left(1+\Delta\mb\right)\,.
\end{equation}
Since the numerical value of $m_b$ is fixed from experiment, equation
\eq{eq:massYukawa} induces a change in the effective Yukawa coupling. This
affects not only the CP-even neutral Higgs field, but the whole Higgs
multiplet, with phenomenological consequences for the charged Higgs particle.
In particular, eq.~\eq{eq:massYukawa} modifies the Yukawa coupling of the
charged Higgs to top and bottom quarks as follows:
\begin{eqnarray}
  \label{eq:effcoupling} 
   h_b\sin\beta \; = \; \frac{\mb}{v} \tb & \longrightarrow  &
   h_b \; = \; \frac{\mb}{v}\,\frac{1}{1+\Delta\mb}\tb,
\end{eqnarray}
where $v=\sqrt{v_1^2+v_2^2}\simeq 174\,$GeV. In the last equation we have
assumed a large \tb\ regime.

%%As emphasized in the introduction, 
It turns out that, in the MSSM with large \tb, the dominant supersymmetric
radiative corrections to the Yukawa interactions of the Higgs doublet
$H_1=(H_1^+,H_1^0)$ stem from the relation \eq{eq:effcoupling}. Explicit loop
corrections to the $H_1 \ov{f} f^\prime$ Yukawa coupling are suppressed by at
least one power of \tb.  This remarkable feature has far-reaching
consequences: first in observables involving the coupling $h_b$ of $H_1$ to
bottom quarks the MSSM behaves like a two-Higgs-doublets model. The main
effect of a heavy SUSY spectrum is to modify the coupling strength via $\Delta
\mb$ in eq.~\eq{eq:massYukawa}, which depends on the masses of the
supersymmetric particles. In certain regions of the parameter space a sizeable
enhancement of $h_b$ occurs.  Secondly these dominant corrections encoded in
$\Delta \mb$ are universal. They are not only equal for the neutral and the
charged Higgs bosons, on which we will focus in the following, but they are
also independent of the kinematical configuration.  This means that they
affect the decay rate of a charged Higgs into a top and bottom (anti-) quark
in the same way as the $\ov{t} b H^+$ vertex in a rare $b$-decay amplitude or,
after replacing the top by a charm quark, as Higgs-mediated $b \rightarrow c$
decays. Further the universality property of these \tb-enhanced radiative
corrections allows for a simple inclusion into the Higgs search analysis.

The proper tool to describe such universal effects is an effective lagrangian.
Expanding~\eq{eq:efflag} to include the charged Higgs sector one finds that
the relevant terms in the large \tb\ limit are:
\begin{eqnarray} 
{\cal L} &=&  
        - h_b                  \, \ov{b}_L b_R \, H_1^0
        + h_b           \, V_{tb}\sin\beta\, \ov{t}_L b_R \, H^+  
        - \Delta h_b          \, \ov{b}_L b_R \, H_2^0 
        + \mathrm{h.c.} \label{eq:uli:lag} 
\end{eqnarray}
$\Delta h_b$ is the loop-induced Yukawa coupling associated with the
supersymmetric QCD corrections in fig.~\ref{fig:vSQCD} and similar electroweak
contributions. $ H^+$ is the physical charged Higgs boson. The Higgs mechanism
defines the relation between the bottom mass $m_b$ and the couplings $h_b$ and
$\Delta h_b$ in ${\cal L}$: calculating the tree-level $\ov{b}bH_1^0$ and
one-loop $\ov{b}bH_2^0$ vertices with zero Higgs momentum, and replacing the
Higgs fields by their vacuum expectation values $v_{1,2}$, yields the desired
relation in eqs.~\eq{eq:massYukawa} and \eq{eq:effcoupling}:
\begin{eqnarray} 
\label{Dmbdef}
\Dmb &=& \frac{\Delta h_b}{h_b} \tb \; = \; \Dmb[SQCD] + \Dmb[SEW]\,,
\end{eqnarray} 
which contains the \tb-enhanced radiative corrections. The supersymmetric QCD
corrections of fig.~\ref{fig:vSQCD} read \cite{copw}
\begin{eqnarray}
  \label{eq:dmbSQCD}
  \Dmb[SQCD] &=& \frac{2\aS}{3\pi}\mg 
                        \mu\tb\,I(\msb{1},\msb{2},\mg)\,.
\end{eqnarray}
Here \aS\ is the strong coupling constant and $\mu$ is the mass parameter
coefficient of the $\epsilon_{ij} H_i^1 H_j^2$ term in the superpotential.
The vertex function $I$, which depends on the masses \msb{1,2} of the two
bottom squark mass eigenstates and the gluino mass \mg, reads \cite{copw}
\begin{equation}
  \label{eq:I}
  I(a,b,c)=\frac{1}{(a^2-b^2)(b^2-c^2)(a^2-c^2)}
  \left(a^2b^2\log{\frac{a^2}{b^2}}
        +b^2c^2\log{\frac{b^2}{c^2}}
        +c^2a^2\log{\frac{c^2}{a^2}}\right)\,.
\end{equation}
An interesting limit of eq.~\eq{eq:dmbSQCD} applies when all mass parameters
are of equal size. One has, depending on the sign of $\mu$
\begin{equation}
  \label{eq:dmbSQCDdec}
  \Dmb[SQCD]=\pm\frac{\aS(Q=\MSy)}{3\pi}\tb\,,
\end{equation}
clearly showing that the effect does not vanish for a heavy SUSY spectrum and
can be of ${\cal O}(1)$ for large \tb\ values.

For sizeable values of the trilinear soft SUSY-breaking parameter $A_t$, the
supersymmetric electroweak corrections are dominated by the charged
higgsino-stop contribution, which is proportional to the square of the top
Yukawa coupling, $h_t=\mt/v_2$. 
Wino-sbottom contributions are generally smaller, being proportional to the
square of the $SU(2)_L$ gauge coupling, g, and to the soft SUSY breaking mass
parameter $M_2$. Neglecting the bino effects, which we found to be numerically
irrelevant, these corrections read \cite{m}
% Bino- and wino-sbottom contributions are
% generally smaller, being proportional to the squares of the corresponding
% $U(1)_Y$ and $SU(2)_L$ gauge couplings and soft SUSY-breaking mass 
% parameters, $g'$, $M_1$ for the bino and $g$, $M_2$ for the wino, 
% respectively. These corrections read \cite{m}
\begin{eqnarray}
  \label{eq:dmbSEW}
  \Dmb[SEW] &=&
  \frac{h_t^2}{16\pi^2}\,\mu A_t\tb\,I(\mst{1},\mst{2},\mu)\nonumber\\
  &-& \frac{g^2}{16\pi^2}\,\mu M_2\tb
    \left[\vphantom{\frac{1}{2}}
            \cst[2]\,I(\mst{1},M_2,\mu)
           +\sst[2]\,I(\mst{2},M_2,\mu)\right.\nonumber\\
  &&\phantom{\frac{g^2}{16\pi^2}\,\mu\tb M_2}
+\frac{1}{2}\csb[2]\,I(\msb{1},M_2,\mu)
+\frac{1}{2}\left.\ssb[2]\,I(\msb{2},M_2,\mu)\right] \,.
%%\nonumber\\
%%  &+&\frac{g'{}^2}{32\pi^2}\,\mu M_1\tb\left[
%%                    \ \,2\,I(\msb{1},\msb{2},M_1)\right.\nonumber\\
%%&&\left.\ \ \ \
%%+\,\left(\frac{1}{3}-\ssb[2]\right)\,I(\msb{1},M_1,\mu)
%%+\,\left(\frac{1}{3}-\csb[2]\right)\,I(\msb{2},M_1,\mu)
%% \right].
\end{eqnarray}
% where we have neglected the effects proportional to $g'$, $M_1$, which
% are numerically very small.

When including radiative corrections, one has to specify the
definition of the quark mass \mb\ appearing in the leading order: \mb\
denotes the pole mass corresponding to the on-shell renormalization
scheme, in which the on-shell self-energy is exactly cancelled by the
mass counterterm.

Note that the supersymmetric corrections contained in \Dmb\ enter $h_b$ in
eq.~\eq{eq:effcoupling} as a factor $1/(1+\Dmb)$. To order $\aS$ one is
entitled to expand this factor as $(1-\Dmb)$. In the phenomenologically most
interesting case of a large $|\Dmb|$ of ${\cal O}(1)$, this leads to
disturbingly large numerical ambiguities. Their resolution seems to require
painful higher-order loop calculations, and a large $|\Dmb|$ may even put
perturbation theory into doubt. Yet these \tb-enhanced contributions have the
surprising feature that they are absent in higher orders:

\textit{There are no contributions to \Dmb\ of order
\begin{eqnarray} 
\lt( \aS \frac{\mu}{\MSy} \tb \right)^n   \label{eq:uli:tbn}
\end{eqnarray} 
for $n\geq 2$.}

Here \MSy\ represents a generic mass of the supersymmetric particles. An
analogous result applies to the electroweak corrections. In other words, to
the considered order, \Dmb\ is a \textit{one-loop exact}\ quantity, and the
factor $1/(1+\Dmb)$ contains the corrections to $h_b$ of the form
in~\eq{eq:uli:tbn} to all orders in \aS.

To prove our theorem, consider possible $n$-loop SUSY-QCD contributions to
\Dmb\ proportional to $\tan^n\beta$: the only possible source of additional
factors of \tb\ is the off-diagonal element of the bottom squark mass matrix,
$-\mu \mb \tb$, which can enter the result via the squark masses as
${\msb{2}}^{\!\!\!2}-{\msb{1}}^{\!\!\!2} \simeq 2 \mb \mu \tb $ or through
counterterms to the squark masses. It is easier to track the factors of $\mu
\mb \tb$ by working with ``chiral'' squark eigenstates and assigning these
factors to ``chirality flipping'' two-squark vertices. Thus any extra factor
of \tb\ is necessarily accompanied by a factor of $\mb \mu$. This dimensionful
factor is multiplied with some power of inverse masses stemming from the loop
integrals.  The next step in our reasoning is to show that the loop integrals
always give powers of $1/\MSy$ and can never produce a factor of $1/\mb[n]$.
The appearance of any inverse power of \mb\ in a loop integral would imply a
power-like infrared singularity in the limit $m_b \rightarrow 0$ with gluino
and squark masses held fixed. But the KLN theorem \cite{k} guarantees the
absence of any infrared divergence in all bare diagrams except for those in
which gluons couple to the $b$-quark lines. A two-loop example of the latter
set is shown in fig.~\ref{fig:uli1}.
\begin{figure}[tb]
\begin{minipage}[b]{0.45\textwidth}
  \begin{center}
    \figulia
  \end{center}
\caption{\label{fig:uli1} Two-loop SUSY-QCD diagram containing a
  large logarithm $\log \lt( \MSy/\mb \right)$.} 
\end{minipage}
\hspace{0.05\textwidth}
\begin{minipage}[b]{0.45\textwidth}
  \begin{center}
    \figulib
  \end{center}
\caption{\label{fig:uli2} One-loop diagram derived from the effective
  lagrangian in \eq{eq:uli:lag} corresponding to the diagram in
  fig.~\ref{fig:uli1}. It contains the large logarithm of fig.~\ref{fig:uli1}
  as $\log \mb/Q$. This logarithm is summed to all orders by solving the
  renormalization group for $h_b$ in eq.~\eq{eq:uli:lag}.}
\end{minipage}
\end{figure}
The infrared behaviour of these diagrams can be studied with the help of the
operator product expansion (OPE). The result of the OPE is nothing but the
effective lagrangian in \eq{eq:uli:lag}.  To apply the OPE to our problem we
first have to contract the lines with heavy supersymmetric particles to a
point, i.e.\ we replace the MSSM by an effective theory in which the heavy
SUSY particles are integrated out.  For the case of the diagram in
fig.~\ref{fig:uli1} this yields the diagram in fig.~\ref{fig:uli2}, in which
the loop-induced interaction is represented by the dimension-4 operator
$\ov{b} b H_2^0$. The information on the heavy SUSY masses is contained in the
Wilson coefficient $\Delta h_b$ in eq.~\eq{eq:uli:lag}. The key feature of the
OPE exploited in our proof is the fact that the effective diagram in
fig.~\ref{fig:uli2} and the original diagram in fig.~\ref{fig:uli1} have the
same infrared behaviour. Power counting shows that the diagram of
fig.~\ref{fig:uli2} has dimension zero. It depends only on \mb\ and the
renormalization scale $Q$.  Since $Q$ enters the result logarithmically, the
diagram of fig.~\ref{fig:uli2} depends on \mb\ as $\log \mb/Q$, no power-like
dependence on \mb\ is possible.  This argument ---essentially power
counting--- immediately extends to higher orders.  Terms from diagrams in
which gluons are connected with the $b$-quark line and one of the
SUSY-particle lines in the heavy loop, are either infrared-finite or
suppressed by even one more power of $\mb/\MSy$, because they are represented
in the OPE by operators with dimension higher than 4. Finally there are
diagrams with counterterms. In mass-independent renormalization schemes the
counterterms are polynomial in \mb.  In the on-shell scheme the diagrams with
counterterms can be infrared-divergent for $\mb \rightarrow 0$, but only
logarithmically. In conclusion the loop integrals cannot give factors of
$1/\mb[n]$. Therefore any correction to $\Delta m_b$ of order $\aS[n] \tan^n
\beta$ comes with a suppression factor of $\mb[n]/\MSy[n]$. Higher-order loop
corrections to $\Delta \mb$ are therefore either suppressed by $\mb/\MSy$ or
lack the enhancement factor of \tb, which proves our theorem.

\begin{figure}[tb]
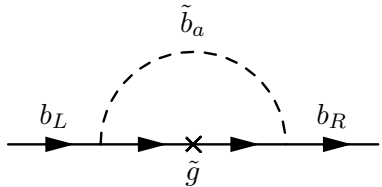

  \begin{center}
    \figmsqcd
  \end{center}
  \caption{\label{fig:mSQCD} One-loop SUSY-QCD diagram contributing to
    \Dmb.}
\end{figure}

So far we have discussed \Dmb\ from the one-loop vertex
function of fig.~\ref{fig:vSQCD} as in \cite{copw}.  A different
viewpoint has been taken e.g.\ in \cite{SQCDHtb}: the
renormalization of the Yukawa coupling $(\mb/v) \tan\beta$ is performed
by adding the mass counterterm to \mb. In the large $\tan\beta$ limit 
and to one-loop order, this amounts to the replacement 
\begin{eqnarray}
  \label{eq:uli:mra}
\frac{\mb}{v} \tb & \longrightarrow  &
   h_b \; = \;   \frac{\mb}{v} \, \lt( 1-\Delta\mb \right)\tb 
\end{eqnarray}
instead of \eq{eq:effcoupling}. This procedure gives the correct
renormalization of the Yukawa coupling in regularization schemes respecting
gauge symmetry \cite{akk}, such as dimensional regularization. The relation to
the Yukawa renormalization using the vertex function in fig.~\ref{fig:vSQCD}
leading to \eq{eq:massYukawa} is provided by a Slavov-Taylor identity
\cite{akk}. 
In general a correction factor related to the anomalous dimension
of the quark mass occurs in \eq{eq:massYukawa}, but the large
\tb -enhanced contributions considered by us are finite and do not
contribute to the anomalous mass dimension. To one-loop order,
eqs.~\eq{eq:uli:mra} and \eq{eq:effcoupling} are equivalent. Yet the
crucial difference here is the point that $-\Dmb$ in
eq.~\eq{eq:uli:mra} stems from the supersymmetric contribution to the
quark self-energy diagram in fig.~\ref{fig:mSQCD}.  While the vertex
diagram has dimension zero, the self-energy diagram has dimension one
and the above proof does not apply.  Indeed, higher-order corrections
to fig.~\ref{fig:mSQCD} do contain corrections of the type in
\eq{eq:uli:tbn}.  In~\ref{sec:Dmb} these corrections are identified
and it is shown that they sum to

%form a geometric series
\begin{equation}     
%%1 - \Delta m_b + \lt( \Delta m_b \right) ^2 -\ldots 
%% &=&  
\frac{1}{1+\Delta\mb}, \nonumber
\end{equation}
so that both approaches lead to the same result \eq{eq:effcoupling}
to all orders in $(\mu/\MSy)\,\aS\tb$.

\subsection{Renormalization group improvement}
\label{ssec:RGi}
The \tb-enhanced supersymmetric corrections discussed so far are not the only
universal corrections.  It is well known that standard QCD corrections to
transitions involving Yukawa couplings contain logarithms $\log \lt( Q/\mb
\right)$, where $Q$ is the characteristic energy scale of the process. For the
decays discussed in sects.~\ref{sec:quantum}--\ref{sec:BRs} one has
$Q=m_{H^+}$ or $Q=\mt$ and $\aS \log \lt( Q/\mb \right)$ is of ${\cal O}(1)$
thereby spoiling ordinary perturbation theory. The summation of the leading
logarithms
\begin{eqnarray}
 \aS[n] \log^n \frac{Q}{\mb},&&\qquad\qquad n=0,1,2\ldots 
\label{eq:uli:ll}
\end{eqnarray}
to all orders in perturbation theory has been performed in \cite{bl}
for the standard QCD corrections to the $\ov{t}_L b_R H^+$ Yukawa
interaction. This summation is effectively performed by evaluating the
running Yukawa coupling $h_b$ at the renormalization scale $Q$.  This amounts
to the use of the running mass at the scale $Q$, $\mb (Q)$, after
expressing $h_b \sin\beta $ in terms of $(\mb/v)\tb$. Hence these large
logarithms are likewise universal, depending only on the energy scale
$Q$ at which the Yukawa coupling is probed, and can also be absorbed
into the effective lagrangian.

The full one-loop QCD corrections to neutral \cite{bl} and charged
\cite{QCDHtb} Higgs decay and top decay \cite{Cz93} also contain
non-logarithmic terms of the order \aS. A consistent use of these one-loop
corrected expressions therefore requires the summation of the next-to-leading
logarithms
\begin{eqnarray}
 \aS[n+1] \log^n \frac{Q}{\mb},&&\qquad\qquad n=0,1,2\ldots 
\label{eq:uli:nll}
\end{eqnarray}
to all orders, because all these terms have the same size as the one-loop
finite terms. Since squarks and gluinos are heavy, leading logarithms of the
type in \eq{eq:uli:ll} are absent in the supersymmetric corrections shown in
fig.~\ref{fig:vSQCD}. It is important to note, however, that this is no longer
true for the next-to-leading logarithms: dressing fig.~\ref{fig:vSQCD} with
$n$ gluons leads to diagrams involving the logarithmic terms
of~\eq{eq:uli:nll}. A two-loop diagram yielding a term of order $ \aS[2] \log
(Q/\mb)$ is shown in fig.~\ref{fig:uli1}.  These subleading logarithms have
escaped attention so far. In the remainder of this section we will address
their proper summation.

In \cite{bl} it has been proved that all leading logarithms occurring in
neutral Higgs decays can be absorbed into the running mass $\mb (Q)$. This
proof is based on the KLN theorem \cite{k} and exploits the fact that there
are only two mass scales, $\mb$ and $Q$, in the loop corrections to neutral
Higgs decay. This reasoning cannot be extended to the next-to-leading
logarithms accompanying the supersymmetric corrections of fig.~\ref{fig:uli1},
where both heavy and light masses occur in the loops. Here we will use the OPE
instead and apply standard renormalization group methods to the effective
coupling in eq.~\eq{eq:uli:lag}. This is not only much more elegant than the
method used in \cite{bl}, it will also show us how to consistently combine the
summation of large logarithms with the all-order result of the \tb-enhanced
terms derived in section~\ref{sec:uli:susy}.

To apply the OPE and the renormalization group one must first employ a
mass-independent renormalization scheme, such as the $\ov{\rm MS}$ scheme
\cite{bbdm}. At the scale $Q=\MSy$ the heavy particles, squarks and gluinos,
are integrated out. The interaction mediated by the loop diagram in
fig.~\ref{fig:vSQCD} is now represented by the effective operator $\ov{b} b
H_2^0$. Its Wilson coefficient equals
\begin{eqnarray} 
-\, \Delta h_b \lt( Q=\MSy \right).
\label{eq:uli:ws}
\end{eqnarray} 
Here and in the following, $\ov{\rm MS}$ quantities are overlined. The
renormalization scale $Q$ is explicitly displayed in~\eq{eq:uli:ws}. Note that
$\Delta h_b$ depends on $Q$ through \aS, \mg\ and the squark masses. The
relation \eq{eq:effcoupling} between $h_b$ and \mb\ is defined at the low
scale $Q=\mb$. Hence we must evolve \eq{eq:uli:ws} down to $Q=\mb$. Since we
encounter the same operator $\ov{b} b $ as in the leading order, the
renormalization group evolution down to $Q=\mb$ is also identical to the
leading-order evolution and just amounts to the use of the running Yukawa
coupling $ \ov{h}_b \lt( Q=\mb \right)$ in the desired relation:
\begin{eqnarray}
   \ov{h}_b \lt(Q=\mb \right) & = &   
 \frac{\ov{m}_b \lt( Q=\mb \right)}{v} \, 
 \frac{1}{1+\Delta\mb \lt( Q=\MSy \right)}\,\tb .
\label{eq:uli:wb}
\end{eqnarray}
Notice that $\Delta\mb$ is evaluated at the high scale $Q=\MSy$: the heavy
particles `freeze out' at the heavy scale $Q=\MSy$ and the strong coupling
\aS\ in $\Delta\mb$ likewise enters the result at this scale. This can be
intuitively understood, as the loop momenta in fig.~\ref{fig:vSQCD} probe the
strong coupling at typical scales of order $\MSy$. Further any renormalization
group running below $Q=\MSy$ is done with the standard model result for
$\beta$-functions and anomalous dimensions. Since the QCD contributions to the
anomalous dimensions of $\ov{h}_b$ and $\ov{m}_b$ are the same, $\ov{h}_b$ at
an arbitrary scale $Q$ is given by
\begin{eqnarray}
   \ov{h}_b \lt(Q\right) & = &   
 \frac{\ov{m}_b \lt( Q \right)}{v} \, 
 \frac{1}{1+\Delta\mb \lt( \MSy \right) }\,\tb .
\label{eq:uli:wq}
\end{eqnarray}
If one expands $\ov{h}_b(\MSy)$ around $\ov{h}_b (\mb) $ to order \aS[2], one
reproduces the large logarithm of the form $\log(\MSy/\mb)$ contained in the
diagram of fig.~\ref{fig:uli1}. The running mass must be evaluated with the
next-to-leading order formula:
\begin{eqnarray}
\ov{m}_b \lt(Q \right) &=& 
   U_6 \lt( Q,\ov{m}_t \right) \cdot U_5 \lt( \ov{m}_t, \ov{m}_b \right) 
 \cdot \ov{m}_b \lt( \ov{m}_b \right) \label{eq:uli:mass},
\end{eqnarray}
where we have assumed that there are no other coloured particles with masses
between $Q$ and $\mt$. The evolution factor $U_f$ reads
\begin{eqnarray}
U_f \lt(Q_2, Q_1 \right)&=& 
    \lt( \frac{\aS \lt( Q_2 \right) }{\aS \lt( Q_1 \right) }\right)^{d^{(f)}} 
    \lt[ 1 + \frac{\aS \lt( Q_1 \right) - \aS \lt( Q_2 \right)}{4 \pi} 
 J^{(f)} \right]    ,\no \\[1mm] 
d^{(f)}  &=& \frac{12}{33 - 2\,f},\no \\[1mm] 
J^{(f)}  &=& - \frac{ 8982 - 504\, f + 40\, f^2}{3\, (33 - 2\,f)^2}.
\label{eq:uli:dj}
\end{eqnarray}
Here $f$ is the number of active quark flavours. For $Q\leq m_t$ one
must replace $ U_6 \lt( Q,\ov{m}_t \right) \cdot U_5 \lt( \ov{m}_t,
\ov{m}_b \right)$ by $ U_5 \lt( Q,\ov{m}_b \right)$ in eq.~\eq{eq:uli:mass}.
$J^{(f)}$ depends on the renormalization scheme, the result in
eq.~\eq{eq:uli:dj} is specific to the $\ov{\rm MS}$ scheme. The $b$-quark
mass in this scheme is accurately known from $\Upsilon(1S)$
spectroscopy and momenta of the $b\ov{b}$ production cross section
\cite{bs}:
\begin{eqnarray}
\ov{m}_b \lt( \ov{m}_b \right) &=& ( 4.25 \pm 0.08 ) \, \mbox{GeV} 
 \label{eq:uli:mb}. 
\end{eqnarray}
Physical observables such as the $H^+$ and top decay rates discussed in
sections~\ref{sec:quantum}-\ref{sec:BRs} are scheme independent to the
calculated order.  Passing to a different renormalization scheme would change
$J^{(f)}$, but in eq.~\eq{eq:uli:mass} the change in $\aS (\ov{m}_b) \,
J^{(5)}$ is compensated by a corresponding change in the numerical value of
$\ov{m}_b \lt( \ov{m}_b \right)$. Likewise the scheme dependence in $\aS (Q)
\, J^{(6)}$ is compensated by the one-loop standard QCD corrections
\cite{QCDHtb,Cz93} to the decay rates. This concludes the discussion of the
universal renormalization group effects. A discussion of additional aspects
specific to the decay rates $\Gamma(t\rightarrow b\,H^+)$ and
$\Gamma(H^+\rightarrow t\,\bar{b})$ can be found in~\ref{sec:uli:app}.

Finally we arrive at the desired effective lagrangian for large \tb:
\begin{eqnarray}
{\cal L} &=&  \frac{g}{2 M_W} \, \frac{\ov{m}_b(Q)}{1+\Dmb} 
     \lt[\vphantom{\lt(\Dmb\,+\frac{\cos\alpha}{\sin\beta}\right)}
         \phantom{\,+}       \, \tb A \, i\,\ov{b}\gamma_5 b  (Q) \right.\nn
&&
     \phantom{\,\,\frac{g}{2 M_W} \, \frac{\ov{m}_b(Q)}{1+\Dmb}}
         +\sqrt{2} \, V_{tb}   \, \tb\, H^+ \,    \ov{t}_L       b_R(Q)
         +\sqrt{2} \, V_{tb}^* \, \tb\, H^- \,    \ov{b}_R       t_L(Q) \nn
&&
     \phantom{\,\,\frac{g}{2 M_W} \, \frac{\ov{m}_b(Q)}{1+\Dmb}}
+\lt(\frac{\sin\alpha}{\cos\beta}-\Dmb\,\frac{\cos\alpha}{\sin\beta}\right) 
                                           h\,\ov{b} b(Q)          \nn
&& 
 \lt.\phantom{\,\,\frac{g}{2 M_W} \, \frac{\ov{m}_b(Q)}{1+\Dmb}}
-\lt(\frac{\cos\alpha}{\cos\beta}+\Dmb\,\frac{\sin\alpha}{\sin\beta}\right) 
                                           H\,\ov{b} b(Q)  
         \,\,\right]\,, \label{eq:uli:lag2}
\end{eqnarray}
where the renormalization scale $Q$ entering $\ov{m}_b$ and the
renormalization constants of the quark bilinears are explicitly shown. In
equation \eq{eq:uli:lag2} we have expressed ${\cal L}$ in terms of the
physical Higgs fields $H,h,A$ and $H^+$ and traded $v$ for the $W$ mass and
the SU(2) gauge coupling $g$. We have used the standard convention
\cite{hhg,Ca98} for these fields and the $h$--$H$ mixing angle $\alpha$.  For
completeness also the coupling of the CP-odd Higgs boson $A$ has been
included. The phenomenology of the neutral Higgs bosons in the large \tb\ 
regime has been studied in detail in \cite{Ca98}.

%For $M_A \gg M_h$ one has $\cos\alpha \simeq \sin\beta$ and $\sin \alpha
%\simeq -\cos\beta$ \cite{hhg,Ca98}.

The effective lagrangian in eq.~\eq{eq:uli:lag2} describes the $A \,
\ov{b}\gamma_5 b$ and $H^+ \, \ov{t}_L b_R$ interactions correctly for large
\tb, irrespective of the mass hierarchy between \MSy\ and \mH. Even if
$\MSy\approx \mH$, the supersymmetric loop form factors of these interactions
are suppressed by one power of \tb\ with respect to the terms described by
${\cal L}$. On the contrary, this is no longer true for the $H\,\ov{b}_L b_R$
and $h\,\ov{b}_L b_R$ form factors \cite{Bo99}. For these couplings ${\cal L}$
is only correct in the limit $\MSy[2] \gg M_A^2$.

\section{Quantum corrections to $\Gamma(t\rightarrow b\,H^+)$, 
$\Gamma(H^+\rightarrow t\,\bar{b})$}
\label{sec:quantum}

The tree-level partial widths read
\begin{eqnarray}
  \label{eq:htreewidth}
  \Gamma^{tree}(\tbH)
  &=&\frac{g^2}{64\pi\mw[2]}|V_{tb}|^2
      \mt[3]\,\lambda^{1/2}\left(1,\qh,\qb\right)\times\nonumber\\
  &&\hspace{1.5cm}\left[\,\left(1-\qh+\qb\right)
      \left(\ctb[2]+\qb\tb[2]\right)+4\,\qb\,\right]\,,\\
  \label{eq:ttreewidth}
  \Gamma^{tree}(\Htb)
  &=&\frac{g^2\,N_c}{32\pi\mw[2]}|V_{tb}|^2
      \mH[3]\,\lambda^{1/2}\left(1,\rt,\rb\right)\times\nonumber\\
  &&\hspace{1.5cm}\left[\,\left(1-\rt-\rb\right)
      \left(\rt\ctb[2]+\rb\tb[2]\right)-4\,\rt\rb\,\right],
\end{eqnarray}
where we have defined the ratios $q_{b,H^+}=m_{b,H^+}^2/\mt[2]$,
$r_{b,t}=m_{b,t}^2/\mH[2]$ and the $\lambda^{1/2}$ term is a kinematic factor
\[
\lambda\left(1,x,y\right)=1+x^2+y^2-2\,(x+y+x\,y)\,.\]

From now on, we shall assume $|V_{tb}|\simeq 1$ and neglect light fermion
generations. For values of the parameter $\tb\gsim 15$ (the inflexion point
being given by $\tb\gsim\sqrt{\mt/\mb}\sim 7$) virtual quantum effects are
largely dominated by the corrections to the right-handed bottom Yukawa
coupling. In that limit the tree-level widths reduce to
\begin{eqnarray}
  \label{eq:tRtreewidth}
  \Gamma^{tree}(\tbH)&=&\frac{g^2\mt[3]}{64\pi\mw[2]}
\,\left(1-\qh\right)^2
      \,\qb\tb[2]\,,\\  %%\left(1+2\,F_R^*\right)\\
  \label{eq:hRtreewidth}
  \Gamma^{tree}(\Htb)&=&\frac{g^2\,N_c}{32\pi\mw[2]}\mH[3]
\,\left(1-\rt\right)^2
\,\rb\tb[2]\,,       %%\left(1+2\,F_R\right)\,,
\end{eqnarray}
in which we have also taken into account the smallness of \mb\ as compared to
\mt, \mH.

\subsection{Standard QCD correction}
\label{sec:QCD}

As we have proved in~\ref{sec:uli:app} applying the OPE, both leading and
subleading $\log(Q/\mb)$ logarithms in the \tbH\ and \Htb\ renormalized decay
widths can be resummed by using the running, \aS[2] corrected, bottom mass in
the zeroth-order expressions. The one-loop finite QCD terms, though, are also
sizeable, and have to be taken into account. In this section we derive
improved expressions for the QCD-corrected decay rates, including both kind of
effects, for large \tb\ values.

The one-loop QCD-corrected expressions for the $t$ ($H^+$) decay rates
\cite{QCDHtb,Dj95,QCDtbH,Cz93} can be greatly simplified after expanding them
in a series in powers of \rb\ (\qb) and retaining only the first-order term.
As we are mainly interested in the region of large \tb, we will provide
formulae valid for those values of \tb, for which eqs.~(\ref{eq:tRtreewidth}),
(\ref{eq:hRtreewidth}) apply. An explicit evaluation of the departure from
this approximation for the one-loop result will be done in
section~\ref{sec:Widths}.

In the $H^+$ case we perform a simultaneous expansion in powers of \rb\ and
\rt. Retaining terms up to \rt[3] and considering the logarithmic factors to
be of ${\cal O}(1)$, the resulting approximation to the one-loop formula is
\begin{eqnarray}
  \label{eq:approxhQCD1loop}%%{eq:approxFRQCD}
  \Gamma_{QCD}^{app}(\Htb)&=&
\frac{g^2N_c}{32\pi\mw[2]}\,\mH[3]\left(1-\rt\right)^2\,\rb\tb[2]\,
            \times\nonumber\\[1ex]
\lefteqn{\hspace{-3cm}
\left\{\,1+\frac{\aS}{\pi} 
\left[\,\vphantom{\frac{1^2}{1_2}}3+6\,\rt+\rt[2]
         -\frac{16}{27}\,\rt[3]+2\,\log(\rb)\right.%\right.
%\left.
+\left(-4\,\rt-\frac{10}{3}\,\rt[2]-\frac{40}{9}\,\rt[3]\,\right)
           \log(\rt)\,\left.\vphantom{\frac{1}{2}}
\vphantom{\frac{1^2}{1_2}}\right]\right\}.}
\end{eqnarray}
As can be seen from the above equation, there is no need to do the resummation
of the $\log\rt$ logarithms, as they are either small when $\rt$ is close to~1
or suppressed by at least a power of \rt\ when it is small.

In the limit of very small \rt, eq.~\eq{eq:approxhQCD1loop} reduces to
\begin{equation}
  \label{eq:limQCD}
  \Gamma_{QCD}(\Htb)=
\left[\,1+\frac{\aS}{\pi}\left(\,3+2\log\rb\,\right)\right]
  \,\Gamma^{(0)}(\Htb)\,,
\end{equation}
where we have introduced the quantity $\Gamma^{(0)}$, which is formally
identical to $\Gamma^{tree}$ but has as input parameters the on-shell
renormalized ones. The finite part in eq.~\eq{eq:limQCD}, $3\aS/\pi$, stands
for a correction of about $+10\%$ (for $\aS\simeq 0.1$), whereas the full
correction is large and negative, due to the much bigger logarithmic term.

For the \tbH\ decay, the expansion in \qb\ reads
\begin{eqnarray}
  \label{eq:approxtQCD1loop}
  \Gamma_{QCD}^{app}(\tbH)&\simeq&
  \frac{g^2\mt[3]}{64\pi\mw[2]}\,
  \left(1-\qh\right)^2\,\qb\tb[2]\,\left\{\,1+\frac{4\aS}{3\pi}
             \,\right.\times\nonumber\\
\lefteqn{\hspace{-1cm}\left[
\,\frac{9}{4}-\frac{2\,\pi^2}{3}+\frac{3}{2}\,\log\qb-\frac{\qh}{1-\qh}\log\qh
+\frac{2-5\,\qh}{2\,\qh}\,\log(1-\qh)\right.}\nonumber\\
\lefteqn{\hspace{2cm}
\left.\left.+\log\qh\log(1-\qh)+2\,\mathrm{Re\,Li_2}(1-\qh)
\vphantom{\frac{\pi^2}{4}}\right]\right\}.}
\end{eqnarray}

In the limit $\qh\rightarrow 1$, the ratio $\Gamma_{QCD}/\Gamma^{(0)}$ becomes
infinite and perturbation theory breaks down, as the $b$-quark moves too
slowly in the top rest frame. Nevertheless, the correction goes to zero due to
the presence of the kinematic suppression factor.

At this point we are ready to incorporate the resummation of the leading and
next-to-leading \qb, \rb\ logarithms, as explained in section~\ref{ssec:RGi},
which amounts to replacing \mb\ in eqs.~\eq{eq:approxhQCD1loop} and
\eq{eq:approxtQCD1loop} by the running bottom mass at the proper
scale.\footnote{We refer the reader to~\ref{sec:uli:app} for a proof of that
  statement.}  The one-loop QCD-corrected widths are then, in the large
$\tan\!\beta$ limit and including renormalization group effects up to
next-to-leading order, given by the following improved ({\em imp}) formulae
\begin{eqnarray}
  \label{eq:imptQCD1loop}
  \Gamma_{QCD}^{imp}(\tbH)&=&
\frac{g^2}{64\pi\mw[2]}\,\mt\,\left(1-\qh\right)^2
       \,\omb[2](\mt[2])\tb[2]\times\nonumber\\
\lefteqn{\hspace{-1.5cm}\left\{\,1+\frac{\oaS(\mt[2])}{\pi}
\left[\,7-\frac{8\pi^2}{9}
-2\,\log(1-\qh)+2\,(1-\qh)\right.\right.}\nonumber\\
&&\left.\left.+\left(\frac{4}{9}
       +\frac{2}{3}\log(1-\qh)\right)(1-\qh)^2\,
\vphantom{\frac{\qb}{1-\qh}}\right]\right\}\,,
\end{eqnarray}
\begin{eqnarray}
  \label{eq:imphQCD1loop}
  \Gamma_{QCD}^{imp}(\Htb)&=&
\frac{g^2N_c}{32\pi\mw[2]}\,\mH\left(1-\rt\right)^2
       \,\omb[2](\mH[2])\tb[2]\times\nonumber\\[1ex]
\lefteqn{\hspace{-1in}
\left\{\,1+\frac{\oaS(\mH[2])}{\pi}\left[\frac{17}{3}
          +6\,\rt+\rt[2]-\frac{16}{27}\,\rt[3]\,
+\left(-4\,\rt-\frac{10}{3}\,\rt[2]-\frac{40}{9}\,\rt[3]\,\right)\log(\rt)
\,\right]\right\},}
\end{eqnarray}
where $\oaS(Q^2)$ is the \MS-scheme running coupling constant and $\omb(Q^2)$
the \MS\ running mass expressed in terms of the bottom pole mass. 

Finite parts in $\Gamma^{imp}$ and $\Gamma^{app}$ differ (see e.g. the $17/3$
in eq.~\eq{eq:imphQCD1loop} and the $3$ in eq.~\eq{eq:approxhQCD1loop}). There
is an implicit scheme conversion in going from eqs.~\eq{eq:approxhQCD1loop},
\eq{eq:approxtQCD1loop} to eqs.~\eq{eq:imptQCD1loop}, \eq{eq:imphQCD1loop}:
the bottom pole mass has been replaced for the running \MS\ mass in the
prefactor and the $\log(\rb)$ has been absorbed into \omb. Notice that the
non-logarithmic terms of $\Gamma_{QCD}^{app}$ have been explicitly included in
$\Gamma_{QCD}^{imp}$, as they are not accounted for by the renormalization
group resummation techniques.

\subsection{Supersymmetric corrections}
\label{sec:SUSY}

The effective lagrangian prediction for the SUSY-QCD and SUSY-EW corrected
decay rates can be read from eq.~\eq{eq:uli:lag2}. No \tb-enhanced vertex
corrections contribute to the matching and the result is obtained by simply
inserting the effective coupling, eq.~(\ref{eq:effcoupling}), into the
zeroth-order width
\begin{equation}
  \label{eq:dmbineffwidth}
  \Gamma_{SUSY}^{eff}=\frac{1}{\left(1+\Dmb\right)^2}\,\Gamma^{(0)}\,.
\end{equation}

We want to compare eq.~\eq{eq:dmbineffwidth} with the diagrammatic on-shell
expressions for the one-loop SUSY-QCD and SUSY-EW corrected \tbH, \Htb\ 
partial widths \cite{SQCDHtb,Gu95,Co96}, which we will denote by
$\Gamma^{1-loop}_{SUSY}$. For large \tb\ values, the only sizeable diagrams
are those that contribute to the scalar part of the bottom quark self-energy,
entering the computation through the mass counterterm. For the SUSY-QCD
corrections, the diagram that matters is shown in fig.~\ref{fig:mSQCD}. By
simple power counting one can realize that it is finite. Moreover, neglecting
${\cal O}(\mb[2]/\MSy[2])$ contributions, its value is essentially given by
that of the three-point diagram in eq.~(\ref{eq:dmbSQCD}): \Dmb[SQCD].
 
Similarly, the diagram relevant to the SUSY-EW corrections is a two-point one
with a chargino (neutralino) and a stop (sbottom) inside the loop. As for the
SUSY-QCD case, it is finite, and its value can be approximated by the
corresponding three-point diagram where an extra $H_2^0$ leg is attached to
the scalar line. Its contribution is thus given by \Dmb[SEW] in
eq.~(\ref{eq:dmbSEW}).

Collecting the results from eqs.~\eq{eq:dmbSQCD} and \eq{eq:dmbSEW} via
eq.~\eq{Dmbdef}, the one-loop SUSY corrected decay rates can be cast into the
formula
\begin{equation}
  \label{eq:dmbinwidth}
  \Gamma_{SUSY}^{1-loop}=
  \left(\vphantom{v^2}1-2\,\Dmb\right)\,\Gamma^{(0)}+\Delta\Gamma_{SUSY}.
\end{equation}
The term $\Delta\Gamma_{SUSY}$, which contains non-universal and
\tb-suppressed contributions to the decay, is very small provided \tb\ is
large, as we have numerically checked.

Both prescriptions, eqs.~(\ref{eq:dmbinwidth}) and (\ref{eq:dmbineffwidth}),
are equivalent at first order in perturbation theory (PT) and consequently do
not differ significantly when the corrections are small. In general, though,
\Dmb\ can be a quantity of ${\cal O}(1)$ for large enough \tb\ values, in
which case eq.~(\ref{eq:dmbineffwidth}) is preferred as it correctly encodes
all higher-order \Dmb\ effects (see the discussion in
section~\ref{sec:uli:susy} and in~\ref{sec:uli:app}).

%%%%%%%%%%%%%%%%%%%%
\subsection{Full MSSM renormalization group improved correction}
\label{sec:hMSSM}

In section~\ref{ssec:RGi}, we saw how the effective lagrangian
\eq{eq:uli:lag2} accounts for the higher-order \tb-enhanced SUSY quantum
corrections and also for the leading and next-to-leading QCD logarithms,
including those in diagrams like fig.~\ref{fig:uli1}. We define the improved
values for the decay rates of the two processes under study in the MSSM as
\[
  \Gamma_{MSSM}^{imp}=\Gamma_{QCD}^{imp}\,\frac{1}{\left(1+\Dmb\right)^2}
  +\Delta\Gamma_{SUSY}\,,
\]
which also incorporates the one-loop finite QCD effects. Neglecting the small
\tb-suppressed $\Delta\Gamma_{SUSY}$ effect, one has
\begin{eqnarray}
  \label{eq:hMSSMimp}
  \Gamma_{MSSM}^{imp}(\Htb)&=&\frac{g^2N_c}{32\pi\mw[2]}\,\mH
  \,\left(1-\rt\right)^2
  \frac{\omb[2](\mH[2])}{\left(1+\Dmb\right)^2}\,\tb[2]
                                    \times\nonumber\\[1ex]
\lefteqn{\hspace{-2.5cm}\left\{\,1+\frac{\oaS(\mH[2])}{\pi}
\left[\frac{17}{3}
      +6\,\rt+\rt[2]-\frac{16}{27}\,\rt[3]
+\left(-4\,\rt-\frac{10}{3}\,\rt[2]-\frac{40}{9}\,\rt[3]\,\right)\log(\rt)
\,\right]\right\},}
\end{eqnarray}
\begin{eqnarray}
  \label{eq:tMSSMimp}
  \Gamma_{MSSM}^{imp}(\tbH)&=&
     \frac{g^2}{64\pi\mw[2]}\,\mt\,\left(1-\qh\right)^2
  \frac{\omb[2](\mt[2])}{\left(1+\Dmb\right)^2}\,\tb[2]\times\nonumber\\[0.5ex]
\lefteqn{\hspace{-2.5cm}\left\{\,1+\frac{\oaS(\mt[2])}{\pi}
\left[\,7-\frac{8\pi^2}{9}
-2\,\log(1-\qh)+2\,(1-\qh)\right.\right.}\nonumber\\
\lefteqn{\hspace{3.5cm}\left.\left.+\left(\frac{4}{9}
       +\frac{2}{3}\log(1-\qh)\right)(1-\qh)^2\,
\vphantom{\frac{\qb}{1-\qh}}\right]\right\}\,.}
\end{eqnarray}

The above formulae contained all the improvements discussed in this article.
In order to compare them to the diagrammatic one-loop MSSM results, we introduce
$\Gamma_{MSSM}^{1-loop}$
\begin{equation}
  \label{eq:1loopMSSMwidth}
  \Gamma_{MSSM}^{1-loop}=\Gamma_{QCD}^{imp}\,
        \frac{\Gamma_{SUSY}^{1-loop}}{\Gamma^{(0)}}\,,
\end{equation}
which only differs from $\Gamma_{MSSM}^{imp}$ in that no resummation of the
SUSY-QCD, SUSY-EW corrections is performed. Comparing
$\Gamma_{MSSM}^{1-loop}$, $\Gamma_{MSSM}^{imp}$ one can assess the size of the
higher-order \tb-enhanced effects.

\section{Results on the decay rates}
\label{sec:Widths}

Although the \tbH\ and \Htb\ decays are mutually exclusive, in the effective
2HDM lagrangian we constructed in section~\ref{sec:efflag}, the supersymmetric
corrections to both observables are encoded in the same effective coupling.
Therefore, we prefer to present the study of these corrections simultaneously,
stressing the points they have in common.

To quantify the importance of the quantum corrections we introduce the
relative correction to the width $\delta$, defined as
\begin{equation}
  \label{eq:ddef}
  \delta\Gamma_x=\frac{\Gamma_x-\Gamma^{tree}}{\Gamma^{tree}}\,.
\end{equation}

\subsection{One-loop vs. NLO-improved QCD corrections}
\label{sec:QCDresult}

Figures~\ref{fig:dtqcd} and \ref{fig:dhqcd} analyse the gluonic corrections to
the \tbH\ and \Htb\ decay rate, showing their dependence on the mass of the
charged Higgs boson for $\tb=10$ and $30$. The dotted lines represent the
relative shifts, \dqcd[1-loop] (\ref{eq:ddef}), produced by the one-loop QCD
corrections, $\Gamma_{QCD}^{1-loop}$, which have been computed using the
formulae in refs.~\cite{Dj95,Cz93}.

In the limit of large \tb\ and small \qb, \rb, the above one-loop results
admit simpler approximate expressions, which we have derived in
eqs.~(\ref{eq:approxhQCD1loop}), (\ref{eq:approxtQCD1loop}). These
approximations have a lower bound of validity, which can be roughly set at
$\tb=10$. In this paper we will not consider values of \tb\ smaller than 10.
Inserting the expansions~(\ref{eq:approxhQCD1loop}),
(\ref{eq:approxtQCD1loop}) into eq.~\eq{eq:ddef}, one obtains the
corresponding relative shifts \dqcd[app] for the \tbH\ and \Htb\ partial
widths, which are plotted using the dashed lines in figs.~\ref{fig:dtqcd}
and~\ref{fig:dhqcd} respectively.

For $\tb=10$, the first term in the \rb\ expansion of the non \tb-enhanced
one-loop QCD corrections to the \tbH\ decay rate stands for a contribution of
about 5\%. For the sake of simplicity, we omitted this term in
eq.~\eq{eq:approxtQCD1loop}, but we have included it when drawing the
\dqcd[app] curve in fig.~\ref{fig:dtqcd}. The extra correction is almost
negligible for the $\tb=30$ curve. In the \Htb\ decay rate,
fig.~\ref{fig:dhqcd}, and for $\tb\ge 10$, eq.~\eq{eq:approxhQCD1loop} is
always extremely close to the one-loop result, and the \dqcd[app] curves are
not shown.

As can be seen in fig.~\ref{fig:dtqcd}, a discrepancy appears between
\dqcd[app] and \dqcd[1-loop] close to the threshold, which can be traced back
to the fact that we dropped the \mb\ kinetic terms in the approximated
formula. Similar problems should be present in the \Htb\ case,
fig.~\ref{fig:dhqcd}, when approaching the threshold, but our plot starts at a
conservative $\mH=250$~GeV value for which the truncated series,
eq.~(\ref{eq:approxhQCD1loop}), with $\rt=0.5$, is still valid. In any case it
makes no sense to try to include higher-order \rt[n] terms because close to
the threshold the perturbative expansion is no longer reliable: the decay
products move slowly in the decay particle's rest frame, and long-distance
non-perturbative effects can significantly modify the perturbative prediction.
Moreover, in this region the branching ratio is very small and therefore the
corresponding decay channel loses its relevance for the charged Higgs
phenomenology.

As was justified in section~\ref{sec:efflag} using the operator product
expansion, the replacement of the renormalized bottom mass and strong coupling
by their running two-loop \MS\ values correctly resums leading and
next-to-leading \rb, \qb\ logarithms. In eqs.~(\ref{eq:imptQCD1loop}),
(\ref{eq:imphQCD1loop}) the substitution was explicitly done. In
fig.~\ref{fig:dtqcd} the numerical effect of the improvement corresponds to
the difference between the dashed (\dqcd[app]) and solid (\dqcd[imp]) curves.
For the \Htb\ decay the improvement is essentially given by the difference
between the dotted (\dqcd[1-loop]) and solid (\dqcd[imp]) curves.

Even for moderate \tb\ values around 10, the QCD corrections are larger than
50\%, driven by the big \qb, \rb\ logarithms.  The resummation of the leading
logarithms is mandatory, specially for the $H^+$ decay where $\log\rb$ is
unbounded as \mH\ increases. The effect of the LO and NLO resummation
diminishes the top partial decay rate in about 5\% and the charged Higgs decay
rate in about 15\%.

\epsfig{dtqcd}{Comparison of the QCD contributions to the \tbH\ decay width,
  as a function of \mH, for $\tb=10$ (upper set) and 30 (lower set). The
  dotted line denotes the one-loop result~\cite{Cz93}, the dashed line the
  approximation of eq.~\eq{eq:approxtQCD1loop}, and the solid line the
  NLO-improved one in eq.~\eq{eq:imptQCD1loop}.}{fig:dtqcd}

\epsfig{dhqcd}{Comparison of the QCD contributions to the \Htb\ decay width,
  as a function of \mH, for $\tb=10$ (upper set) and 30 (lower set). The
  dotted line corresponds to the one-loop~\cite{Dj95} correction, and the
  solid line to the NLO-improved result, eq.~\eq{eq:imphQCD1loop}.}{fig:dhqcd}

\subsection{Supersymmetric corrections}
\label{sec:SUSYresult}

Figure~\ref{fig:dhsusy} focuses on the genuine supersymmetric corrections to
the \Htb\ partial width. As they are dominated by the universal \Dmb\ effect,
the results for the \Htb\ plot represent fairly well the effects of the
corrections on \tbH\ too. Curves are shown for two values of the
$\mu$-parameter and for two different sparticle spectra.

In the ``heavy'' spectrum, the gluino and the lightest sbottom and stop have a
common 1~TeV mass. The squarks and gluinos are nearly degenerate and they are
much heavier than the $H^+$ mass, justifying the use of the effective
lagrangian approach. As only $\tb\geq 10$ values are considered, the
approximation consisting in neglecting the non-universal and \tb-suppressed
terms denoted by $\Delta\Gamma_{SUSY}$ in eq.~\eq{eq:dmbinwidth}, which is
represented by the dashed \dsusy[app] curves, fits very well the one-loop
calculation (the latter is not shown in this case). The corresponding
effective lagrangian prediction, eq.~(\ref{eq:dmbineffwidth}), which includes
all \tb-leading terms appearing at higher orders in PT, is represented by the
solid \dsusy[eff] lines.

A second, lighter, spectrum is defined by $\mg=500$~GeV, and the masses of the
lightest sbottom and stop around 200~GeV. The curve labelled \dsusy[1-loop]
corresponds to the full one-loop computation, including all possible gluino,
chargino and neutralino loops.  Even for this light spectrum and for the
chosen set of parameters, \dsusy[app] gives a good estimate of the one-loop
correction. This illustrates the fact that our effective lagrangian ${\cal L}$
in eq.~\eq{eq:uli:lag2} describes the charged Higgs interaction correctly even
if $\MSy < \mH$. It shows that \Dmb\ accounts for most of the effects and we
can trust the validity of the improved result.

Typical values we found for the SUSY correction are 15\%--30\% with the heavy
spectrum and $\sim 40$\% with the light one. In both cases, the results depend
heavily on the $\mu$ and \tb\ parameters, the size of the correction growing
almost linearly with their absolute values. Although not shown in the plots,
the main contribution to \dsusy\ comes from the SUSY-QCD diagrams. Only for a
very large $A_t$ values can the electroweak corrections be comparable.

The \dsusy[eff] curves correspond to the relative correction to the widths as
evaluated using eq.~\eq{eq:dmbineffwidth}, an expression derived from the
effective lagrangian in section~\ref{sec:efflag}. While \dsusy[1-loop],
\dsusy[app] do not include higher-order \Dmb[n] effects (which can be
potentially of ${\cal O}(1)$) these \tb-dominant effects are correctly
resummed to all orders in PT in the expression for \dsusy[eff].

The difference between \dsusy[eff] and \dsusy[app] first appears at order
$(\Dmb)^2$, and is always positive, opposite to the negative standard QCD
corrections, for $\Dmb>-1.5$.\footnote{The comparison is between
  $1/(1+\Dmb)^2$ and $1-2\Dmb$, eq.~(\ref{eq:dmbinwidth}), the approximated
  one-loop result as defined in this paper and in \cite{Co96,Co98}.}
Therefore, for negative (positive) values of \Dmb, that is, positive
(negative) corrections \dsusy[1-loop], the higher-order terms tend to
reinforce (suppress) the correction. As \Dmb\ is mainly given by the SUSY-QCD
contribution, eq.~(\ref{eq:dmbSQCD}), this correlation is seen in association
with the sign of $\mu$.

Just to give some examples, for a negative $\dsusy[1-loop]=30$\% correction,
which corresponds to $\Dmb=-0.15$, the extra higher-order terms contained in
\dsusy[eff] increase the partial width by $8$\%. For $\Dmb=-0.2$, a number
that can be obtained from eq.~(\ref{eq:dmbSQCDdec}) by setting $\tb=20$,
$\aS=0.1$, the difference between $\dsusy[eff]$ and $\dsusy[1-loop]$ is of
order $+16$\%.

The only restriction to the potential size of \dsusy\ is set by the
renormalized bottom Yukawa coupling, which is required to remain perturbative
from the GUT scale to the scale of the corresponding decay. This is guaranteed
in our calculations by demanding $h_b<1.2$ at low energies (see e.g.
\cite{copw}), implying the following combined bound on \tb\ and \Dmb:
\begin{equation}
  \label{eq:tbDmbbound}
  \Dmb>\frac{1}{1.2}\,\frac{g\,\omb(\mt[2])}{\sqrt{2}\mw}\tb-1
       \simeq 0.014\tb-1\,.
\end{equation}
In the above example, with $\tb=20$, the minimum allowed value for \Dmb\ is
$-0.72$. If eq.~(\ref{eq:dmbSQCDdec}) holds for negative $\mu$, and using
$\aS(\MSy)\sim 0.1$, it is found that a maximum allowed correction,
$\dsusy[eff]\gsim+200$\%, is obtained around $\tb=40$.

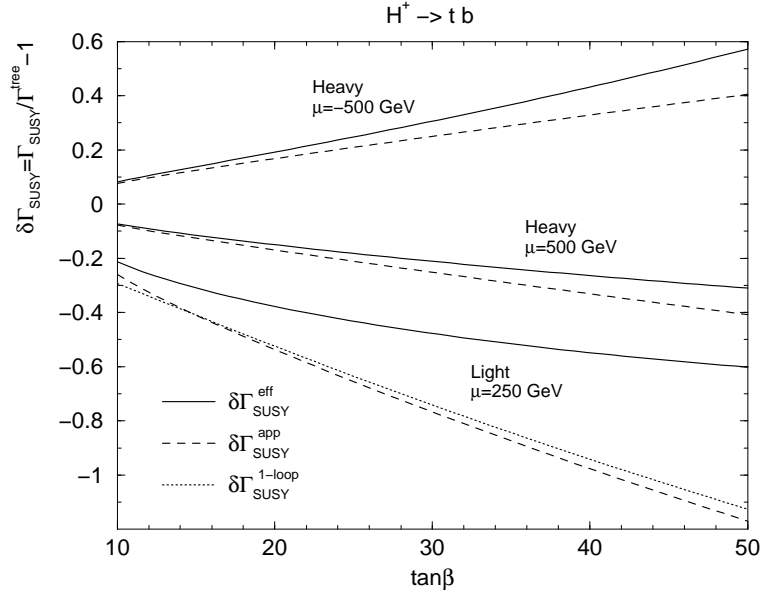
\epsfig{dhsusy}{The SUSY contributions to the \Htb\ partial decay width, as a
  function of \tb, for $\mH=350$~GeV and two values of $\mu$. The dashed lines
  denote the approximation $\Delta\Gamma_{SUSY}=0$ of eq.~\eq{eq:dmbinwidth},
  whereas the solid lines correspond to the effective width,
  eq.~\eq{eq:dmbineffwidth}. For the heavy spectrum one has
  $\mg=m_{\tilde{b}_1}=m_{\tilde{t}_1}=1$~TeV, $\tilde{b}_1$, $\tilde{t}_1$
  being the lightest sbottom and stop respectively. $A_t=500$~GeV, the $\mu$
  values are shown in the plot. For the light spectrum we have set
  $\mg=500$~GeV, $m_{\tilde{b}_1}=250$~GeV and $m_{\tilde{t}_1}=180$~GeV. In
  this case, we also show a dotted curve corresponding to the one-loop result
  of ref.~\cite{Co98}.}{fig:dhsusy}

\subsection{Full MSSM correction}
\label{sec:MSSMresult}

We shall now show the combined effects of the QCD, SUSY-QCD and SUSY-EW
corrections in the partial decay widths under study, starting from
three different sets of curves: \dqcd[imp], i.e. the QCD correction including
the renormalization group resummation of the bottom mass logarithms up to NLO;
\dmssm[1-loop], the full one-loop MSSM contribution as defined in
eq.~(\ref{eq:1loopMSSMwidth}); and the MSSM-improved contribution,
\dmssm[imp], defined in eq. (\ref{eq:tMSSMimp}).

Figure~\ref{fig:dt_msusy} shows the dependence of the relative corrections to
the width $\delta\Gamma(\tbH)$ on the mass scale \MSy, defined as a common
value for the gaugino mass, $M_2$, the gluino mass and the masses of the
lightest stop and sbottom. As we keep the value of $\mu$ fixed, the SUSY
contribution smoothly goes to zero like $\mu/\MSy$ when \MSy\ increases.
Contrarily, if all mass parameters are sent to infinity together, the SUSY
correction tends towards a constant value, determined by
$\Dmb\simeq\pm(\aS/3\pi)\tb$, eq.~(\ref{eq:dmbSQCDdec}). A similar behaviour
occurs for $\delta\Gamma(\Htb)$ with a different renormalized value for \dqcd.

The difference between \dqcd[imp] and \dmssm[imp] is due to the SUSY
corrections, which were already considered in the above section. The mismatch
between \dmssm[1-loop] and \dmssm[imp] is produced by the \tb-enhanced
higher-order effects that are resummed in the latter.
%%% Finally there is an extra effect coming from  

Figure~\ref{fig:dt_tbeta} shows how the full MSSM correction evolves with \tb.
While \dqcd\ has a mild dependence on \tb\ that is almost saturated around
$\tb=20$, the SUSY part gets more and more important as \tb\ increases. One
can see that for the chosen parameters \dsusy\ becomes of ${\cal O}(10\%)$
around \tb=30. For negative values of $\mu$, of ${\cal O}(\MSy)$, and for
sufficiently large \tb\ values, the total correction can be considerably
reduced with respect to the naive QCD prediction. A similar behaviour is found
for $\delta\Gamma(\Htb)$.

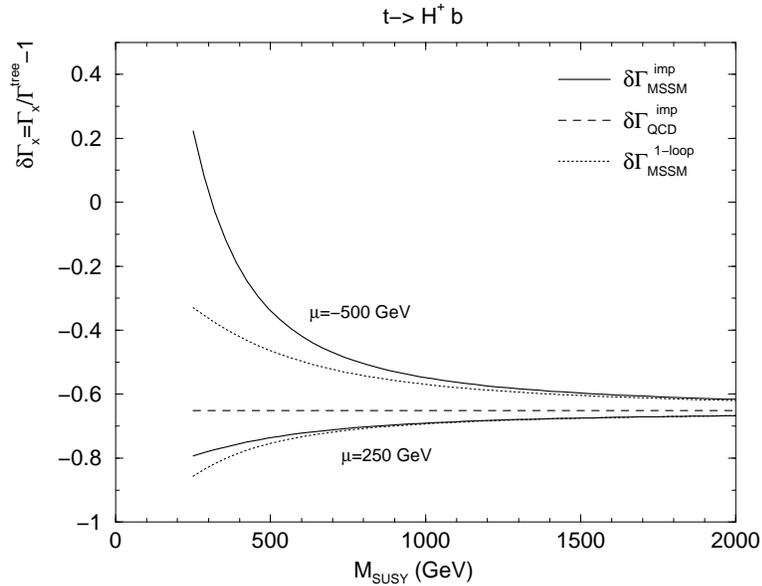
\epsfig{dt_msusy}{Evolution of the corrections to the \tbH\ width, for
  $\mH=125$~GeV and $\tb=30$, as a function of a ``common SUSY mass'',
  $\MSy=M_2=\mg=m_{\tilde{b}_1}=m_{\tilde{t}_1}$, and $A_t=500$~GeV. The
  dashed line corresponds to the QCD-improved width,
  eq.~(\ref{eq:imptQCD1loop}), the dotted line denotes the one-loop MSSM
  result, eq.~(\ref{eq:1loopMSSMwidth}), and the solid line denotes the
  MSSM-improved one, eq.~(\ref{eq:tMSSMimp}).}{fig:dt_msusy}

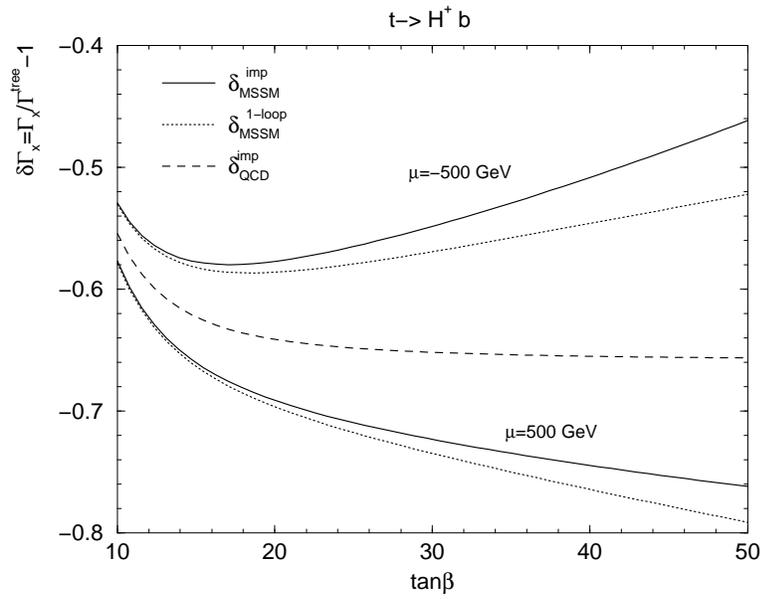
\epsfig{dt_tbeta}{The corrections to the \tbH\ width for $\mH=125$~GeV as a
  function of \tb. The rest of the parameters are those of the heavy spectrum
  in fig.~\protect{\ref{fig:dhsusy}}. The dashed line corresponds to the
  QCD-improved width, eq.~(\ref{eq:imptQCD1loop}), the dotted line denotes the
  one-loop MSSM result, eq.~(\ref{eq:1loopMSSMwidth}), and the solid line
  denotes the MSSM-improved one, eq.~(\ref{eq:tMSSMimp}).}{fig:dt_tbeta}

\clearpage

\section{Results on the branching ratios}
\label{sec:BRs}

Above we have described the effects of the QCD, SUSY-QCD and SUSY-EW
corrections on the decay widths of \tbH\ and \Htb\ as a function of the MSSM
parameter space.  In the case of \tbH, assuming that the only other possible
decay channel is $t\rightarrow b W^+$, we shall present the results on the
${\cal BR}(\tbH)$ and we shall use these computations to exemplify how much
the radiative corrections implemented here can change the actual reach of the
Tevatron collider in the search of $H^+$ in the indirect mode, missing
leptons/dileptons in the $t\rightarrow b W^+$ decay.

Here, the results from the frequentist analysis of D0 indirect $H^+$ searches
\cite{Dhiman} are used to derive constraints on the \tb--\mH\ plane (see e.g.
\cite{TEV} for results on similar indirect $H^+$ searches by the CDF
collaboration).

In fig.~\ref{fig:QCDt_BR} we draw curves of constant ${\cal BR}(\tbH)$ based
on $\Gamma_{QCD}^{imp}(\tbH)$, eq.~\eq{eq:imptQCD1loop}, and including the
one-loop QCD corrections into the computation of $\Gamma(t\rightarrow bW^+)$.
We do not show curves that have a branching ratio smaller than $0.1$ because,
for such regions of parameters, the \tbH\ decay channel has little
phenomenological relevance. The grey area at the bottom-right corner of the
figure is the region excluded by the D0 frequentist analysis data.

The plots in fig.~\ref{fig:QCDt_BR} compare to the plots in
fig.~\ref{fig:MSSMt_BR}.  Here we show curves of constant ${\cal BR}(\tbH)$,
using the MSSM-improved formulae for the partial \tbH\ decay rate,
eq.~\eq{eq:tMSSMimp}. The soft SUSY-breaking masses are chosen to produce a
heavy SUSY spectrum, with $\mg=\mst{1}=\msb{1}=1$~TeV. As in
fig.~\ref{fig:QCDt_BR}, the dark area on the bottom-right corner corresponds
to the experimentally excluded region.

For positive values of \Dmb\ (left plot in fig.~\ref{fig:MSSMt_BR}), both QCD
and SUSY-QCD corrections reduce the tree-level partial width of the \tbH\ 
decay channel, and the bound on the ${\cal BR}$ moves to higher \tb\ values.
In our example plot, with $\mu=500$~GeV, the excluded region starts at
$\tb>100$ and it is not shown. Conversely, for negative \Dmb\ values, the
supersymmetric corrections partly compensate for the QCD reduction of the
width, and the bound is found for lower \tb\ values. This fact can be checked
in the plot on the right of fig.~\ref{fig:MSSMt_BR}, corresponding to
$\mu=-500$~GeV. Values larger than $0.4$ for ${\cal BR}(\tbH)$ are obtained
when $\tb\gsim55$. The experimental bound starts around $\tb=65$, in a region
where $h_b(\mt)>1.2$, which implies that the bottom Yukawa coupling becomes
non-perturbative below the GUT scale \cite{copw}. This fact is denoted in the
plots by changing from solid to dashed line style. The same remark applies for
fig.~\ref{fig:QCDt_BR}.

The \Htb\ branching ratio, which is expected to be tested at the LHC and at
the NLC, is depicted in fig.~\ref{fig:H_BR}. On the left plot, contour lines
of constant ${\cal BR}$ are drawn using the QCD improved width,
eq.~\eq{eq:imphQCD1loop}. Similarly, the right plot shows curves of constant
${\cal BR}(\Htb)$ for the MSSM-improved result, eq.~\eq{eq:hMSSMimp}, with
$\mu=500$~GeV, the rest of SUSY parameters being equal to those of
fig.~\ref{fig:MSSMt_BR}. It has been assumed that no decays of $H^+$ into
pairs of R-odd SUSY particles \cite{Bo98} were possible. This is guaranteed by
the choice of the soft SUSY-breaking masses and by cutting the plots at
$\mH=500$~GeV.

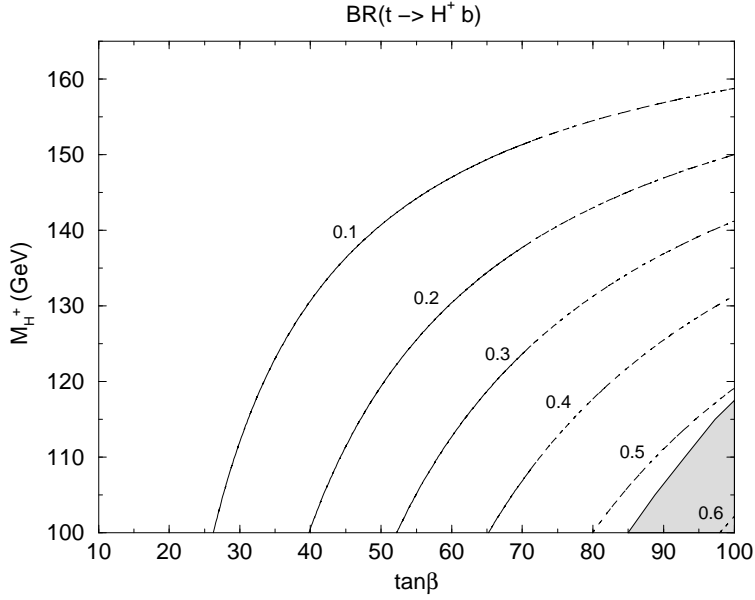
\epsfig{t_bH_QCDi}{Curves of constant branching ratio for the \tbH\ channel.
  The figure shows the QCD-improved, eq.~\eq{eq:imptQCD1loop}, result. The
  transition between the solid and dashed styles occurs when the bottom Yukawa
  coupling crosses the bound $h_b(\mt)<1.2$. As explained in the text, this
  bound guarantees the perturbativity of the Yukawa up to the GUT scale.
  Finally, the shaded area defines the 95\% C.L. exclusion boundary in the
  \tb--\mH\ plane for $\mt=175$~GeV and $\sigma(t\bar{t})=5.5$~pb that can be
  derived from the D0 frequentist analysis in
  ref.~\cite{Dhiman}.}{fig:QCDt_BR}

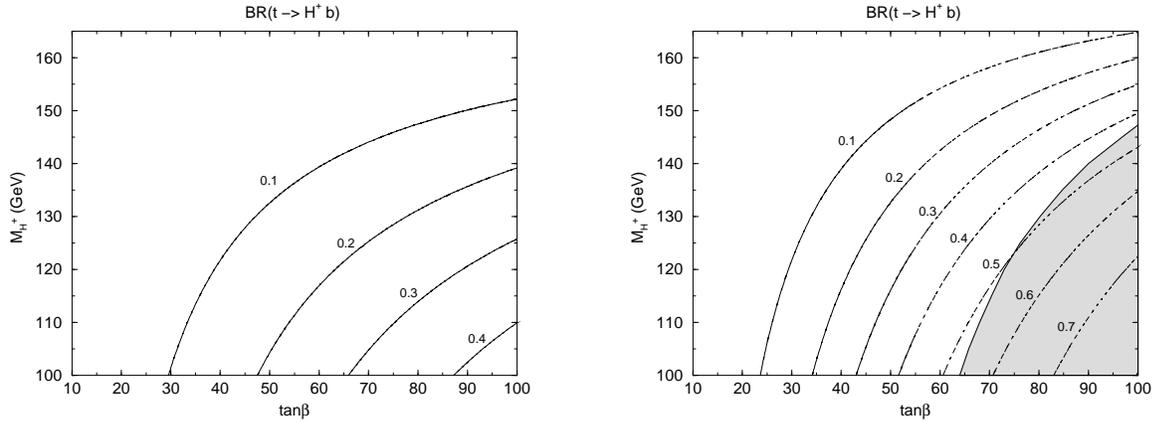
\tepsfig{pt_bH_MSSMi}{mt_bH_MSSMi}{As fig.~\ref{fig:QCDt_BR}, but plotting the
  MSSM-improved result, eq.~\eq{eq:tMSSMimp}, for $\mu=500$~GeV (left plot)
  and $\mu=-500$~GeV (right plot).
%  The right-end of the curves corresponds to the bound $Y_b(1 TeV)=h_b(1
%  TeV)^2/4\pi<0.5$, the minimum required if no new physics is to be found
%  before the TeV scale.
  The rest of relevant SUSY parameters are given by
  $\mg=M_2=\mst{1}=\msb{1}=1$~TeV, $A_t=500$~GeV. In the $\mu=-500$~GeV plot,
  the shaded area is excluded by the D0 frequentist analysis in
  ref.~\cite{Dhiman}.}{fig:MSSMt_BR}

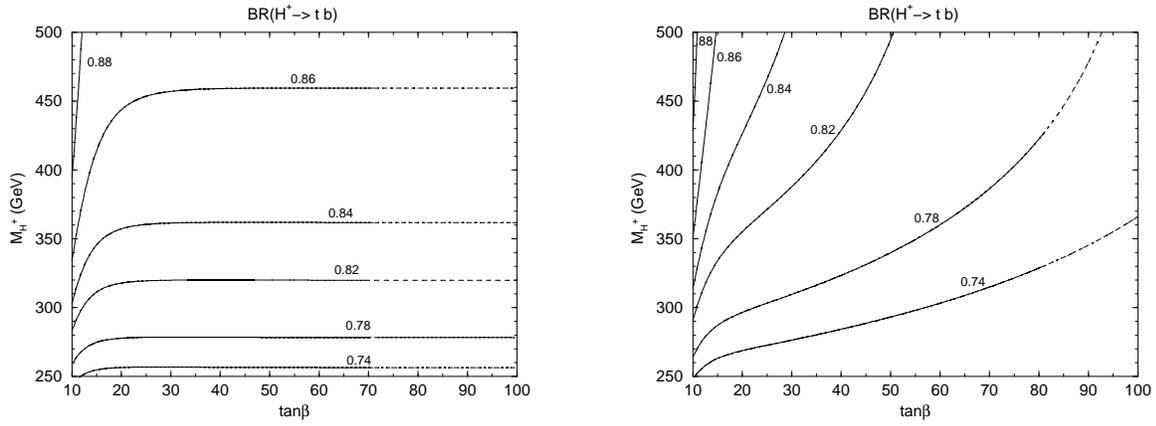
\tepsfig{H_tb_QCDi}{H_tb_MSSMi}{Curves of constant branching ratio for the
  \Htb\ channel. On the left, the QCD-improved values,
  eq.~\eq{eq:imphQCD1loop}, on the right, the MSSM-improved result,
  eq.~\eq{eq:hMSSMimp}. The parameters chosen for these plots are
  $\mu=500$~GeV, $\mg=M_2=\mst{1}=\msb{1}=1$~TeV, $A_t=-500$~GeV.}{fig:H_BR}

\clearpage

\section{Conclusions}
\label{sec:concl}

Using an effective lagrangian description of the MSSM, we have
investigated the virtual supersymmetric effects that modify the
tree-level relation between the bottom Yukawa coupling and the bottom
mass, which are dominant in the large \tb\ regime. Motivated by the
fact that these effects do not vanish for large values of the SUSY
masses and are potentially of ${\cal O}(1)$, we have derived the
expressions for the bottom Yukawa couplings that resum all
higher-order \tb-enhanced quantum effects. These expressions have a
natural interpretation and are easily deduced in the context of the
effective lagrangian formulation. We have also shown that they can be
equivalently deduced in the framework of the full MSSM.

As an interesting application of our results, we have computed the
partial decay rates for the \tbH\ and \Htb\ decay channels, relevant
to supersymmetric charged Higgs searches at present and future
colliders. First we have considered the QCD quantum corrections to
these processes and, applying the OPE, we have performed the
resummation of the leading and next-to-leading logarithms of the form
$\log\,Q/\mb$. Concerning the supersymmetric corrections, we have
compared our results with those of previous diagrammatic one-loop
analyses in the literature and we have shown the numerical relevance
of the resummation of the \tb-enhanced effects derived in this work.
Collecting the above improvements, we have finally computed the
corresponding branching ratios, ${\cal BR}(\tbH)$ and ${\cal
  BR}(\Htb)$. As an example, we have shown, for different sets of the
MSSM parameters, the effect of the quantum corrections in determining
the region of the \mH--\tb\ plane excluded by the D0 indirect searches
for a supersymmetric charged Higgs boson in the decay of the top
quark.

%\nosubsectionnumber
\setcounter{secnumdepth}{0}
\section{Acknowledgements}

We would like to thank D.~Chakraborty for providing us with the frequentist
analysis data on the indirect charged Higgs search at D0. We are also grateful
to L.~Groer for sending us the corresponding data for the direct charged Higgs
search at CDF. U.N.\ thanks M.~Spira and P.M.~Zerwas for a clarifying
discussion on Yukawa coupling renormalization and gauge symmetry. 
% D.G.  acknowledges the financial support from a Marie Curie EC 
% grant TMR-ERBFMBICT983539.
% D.G. acknowledges the support of the European Commission TMR programme 
% grant ERBFMBICT983539.
The work of D.G. was supported by the European Commission TMR programme under
the grant ERBFMBICT 983539.
\setcounter{secnumdepth}{2}

\myappendix

\newpage
\clearpage

\section{The effect of \Dmb\ at all orders}
\label{sec:Dmb}
In this appendix we perform the resummation, in perturbation theory,
of the leading supersymmetric effects contained in \Dmb\ and find
agreement with the effective lagrangian result of sect.~\ref{sec:efflag}.

When relating \tb\ to physical decay rates or to quark masses and
Yukawa couplings one also has to address the question of the proper
treatment of standard QCD corrections. This issue seems to come into
play in the very beginning when one defines the renormalization scheme
and scale for the bottom quark mass, which, for example, enters the
off-diagonal elements of the $b$-squark mass matrix. Here we want to
stress that the issues of large supersymmetric \tb -enhanced
corrections related to the diagram in fig.~\ref{fig:mSQCD} and the
treatment of standard QCD corrections related to gluonic corrections
to the quark self-energy can be treated independently of each other.
In the following we shall concentrate on the \tb-enhanced SUSY-QCD
corrections, induced by supersymmetric particle loop effects. After
performing the resummation of these corrections to all orders in
perturbation theory, one can subsequently include the standard QCD
corrections, whose proper treatment is discussed in
sect.~\ref{sec:efflag}.

% Hence, in the following, we can ignore the gluonic corrections.
% After resumming the supersymmetric \tb -enhanced SUSY corrections to
% all orders one can subsequently switch on the gluonic QCD corrections,
% whose proper treatment is discussed in sect.~\ref{sec:efflag}.

The quantity \Dmb\ is proportional to $(\aS/\pi)(\mu\tb/\MSy)$ and of ${\cal
  O}(1)$ when, simultaneously, $\mu\sim\MSy$ and \tb\ is large. In that case
one should resum its effects to all orders in PT to obtain a reliable
prediction. As was shown in section~\ref{sec:efflag}, the first thing one
should realize is that there are no higher-loop diagrams contributing to the
mass renormalization (nor to the decay rate) of order $\aS[n]\tb[n]$ with
$n>1$. Diagrams with extra $\mu\tb$ insertions are suppressed by powers of
$\mb[2]/\MSy[2]$. This can be easily seen in the effective lagrangian
approach, where such contributions would arise from higher-dimensional
operators with more Higgs boson fields, whose couplings are suppressed by
extra powers of \MSy.

\begin{figure}[ht]
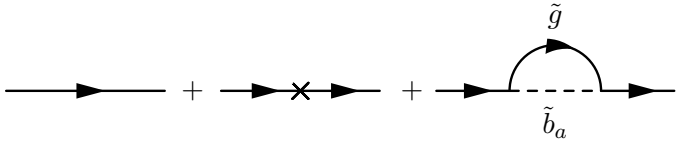

  \begin{center}
\figselfoloop
  \end{center}
  \caption{\label{fig:self1loop} Feynman diagrams contributing to 
    the bottom pole mass up to first order in PT. From left to right, the
    renormalized bottom mass, the bottom mass counterterm and the finite
    one-loop Feynman graph contributions are shown. The dashed line in the
    last diagram denotes a sbottom and the solid line a gluino. The cross
    represents the insertion of the bottom mass counterterm.}
\end{figure}

Different renormalization schemes use different values for the renormalized
bottom Yukawa coupling $h_b$ \cite{bl}. In theories with spontaneous symmetry
breaking, though, there is always a link between the value of $h_b$ and the
physical bottom mass, \mb: the dressed bottom propagator must have a pole for
on-shell external legs, or conversely the inverse propagator must vanish. At
one loop this relation reads, considering only the gluino corrections
\begin{equation}
  \label{eq:mbvshb1loop}
  h_b v_1 + \delta h_b v_1 + h_b v_1\,\Dmb = \mb\,,
\end{equation}
$\delta h_b$ being the counterterm of $h_b$. The l.h.s. of the previous
equation is graphically depicted in fig.~\ref{fig:self1loop}.

We are not displaying the wave function renormalization to avoid an
unnecessary complication of the argument. Note that $v_1$ receives no
one-loop QCD corrections and thus its renormalization only adds
effects suppressed by $\alpha_{EW}/\aS$, which allows us to identify
$\delta h_b v_1$ with $\delta\mb$. Besides, in any renormalization
scheme one has $\mb[R]=h_b v_1$, with \mb[R] and $h_b$
denoting renormalized quantities. Therefore,
\begin{equation}
  \left(h_b+\delta h_b\right)v_1=\mb[R]+\delta\mb\,,
\end{equation}
and one obtains, at first order
\begin{equation}
  \label{eq:mbbare}
  \mb[R]+\delta\mb=\mb-\mb[R]\Dmb.
\end{equation}
The l.h.s. of eq.~(\ref{eq:mbbare}) is just the bare bottom mass, \mb[0].

When evaluated beyond first order, scheme differences appear in the
r.h.s. of eq.~(\ref{eq:mbbare}). In the on-shell scheme, the
renormalization condition being given by $\mb=h_b v_1=\mb[R]$, one
would obtain that the bare bottom mass is equal to
$\mb\left(1-\Dmb\right)$, while in the \MS-scheme, for which
$\delta\omb$ is zero as \Dmb\ is finite, one would have
$\mb/(1+\Dmb)$. Both results are equivalent at first order in
\Dmb, as they should.

\begin{figure}[ht]
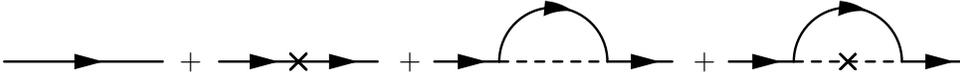

  \begin{center}
\figself
  \end{center}
  \caption{\label{fig:self} Full set of SUSY-QCD dominant diagrams, 
    in the large \tb\ limit, contributing to the bottom pole mass at all
    orders in PT. The first three diagrams are those of
    fig.~\ref{fig:self1loop}. In the fourth one, the cross denotes the
    insertion of the $\delta\tilde{h}_b$ counterterm, and the solid and dashed
    lines denote gluino and sbottom propagators respectively (see
    fig.~\ref{fig:extra}).}
\end{figure}

To proceed with the resummation, we come back to the relation between the
Yukawa and the pole mass. Although no $n$-loop diagrams produce $\aS[n]\tb[n]$
corrections for $n\geq 2$, there is one and only one genuine $n$-th-order
diagram left (see fig.~\ref{fig:self}), which contains the insertion of a
$(n-1)$-loop counterterm into a one-loop diagram. Then, all dominant terms in
the large \tb\ limit, at all orders in PT, are contained in the equation
\begin{equation}
  \label{eq:mbvshb}
  h_b v_1 + \delta h_b v_1 
+ \tilde{h}_b v_1\,\Dmb + \delta\tilde{h}_b v_1\,\Dmb = \mb.
\end{equation}
Beyond tree level, the $\tilde{b}_L\tilde{b}_RH_2^0$ coupling is no longer
equal to $h_b$, so it is denoted by $\tilde{h}_b$, with counterterm
$\delta\tilde{h}_b$.\footnote{The tree-level coupling is in fact $\mb\mu\tb$,
  but again neither $\mu$ nor \tb\ receive QCD corrections at first order.}
This fact was not important in eq.~(\ref{eq:mbvshb1loop}) because we were just
considering the first-order result.

Before proceeding, one technical point in equation~(\ref{eq:mbvshb}) deserves
further clarification. The last term in the l.h.s. corresponds to the true
three-point diagram in fig.~\ref{fig:self}. In the large \tb\ limit, though,
its value, $\delta\tilde{h}_b v_1\Dmb$, coincides with the two-point
contribution, $\tilde{h}_b v_1\Dmb$, after replacing the renormalized coupling
by the counterterm. A derivation of this result is written at the end of this
appendix.

The last step in our argument is to justify the equality
\begin{equation}
  \label{eq:barevsbare}
  h_b + \delta h_b = \tilde{h}_b + \delta\tilde{h}_b\,,
\end{equation}
which can be regarded as the identity of the bare quark and squark Yukawa
couplings, which is guaranteed by the underlying supersymmetry governing the
relations between the bare lagrangian parameters in the
ultraviolet.\footnote{This is true if a regularization method preserving SUSY
  is used, such as dimensional reduction. Deviations from
  eq.~\eq{eq:barevsbare} in the \MS-scheme will be loop-suppressed, not
  affecting the conclusions of this appendix.} No soft SUSY-breaking
dimensionful couplings can induce modifications to eq.~(\ref{eq:barevsbare}),
allowing for the extraction of a common $h_bv_1+\delta h_bv_1$ factor in
(\ref{eq:mbvshb}). At the level of bare couplings one does not need to make
reference to any particular renormalization scheme. Thus, one has
\begin{equation}
  \label{eq:mbvsbare}
  \left(h_b + \delta h_b\right)v_1 = \mb[R] + \delta\mb = \frac{\mb}{1+\Dmb},
\end{equation}
where the r.h.s. is expressed in terms of physical quantities, \Dmb\ being
independent of \mb.

For the rest of this appendix we will derive expressions valid to all
orders in PT in the large \tb\ limit for the $H^+$ and $H,h,A$ dressed
couplings to $t\bar{b}$ and $b\bar{b}$ respectively, recovering the
effective lagrangian results one can find in \cite{Ca98}. The
calculation involves contributions from three-point loop diagrams with
one external on-shell Higgs leg whose momentum we have neglected. In
section~\ref{sec:Widths}, the departure from this assumption for the
$H^+$ and $t$ decay rates has been shown to be small, as the extra
contribution inducing the momentum dependence does not include any
\tb\ enhancement factor. More complete formulae including the momentum
dependence for the decay rates of the neutral Higgs bosons can be
found in ref.~\cite{Bo99}.

Let us start with the simplest case, that of the charged Higgs $H^+$ and of
the pseudoscalar $A$, for which there are no vertex loop diagrams \tb-enhanced
with respect to the tree-level coupling. The relevant Feynman diagrams are
just the tree-level Yukawa and the counterterm. From eq.~\eq{eq:mbvsbare}, the
renormalized decay amplitudes are given by
\begin{eqnarray}
  \label{eq:effHpA0Yuk}
  i\left(h_b+\delta h_b\right)\sin\beta\,H^+\bar{t}P_Rb&=&
  i\,\frac{\mb\tb}{\left(1+\Dmb\right)v}
                        \,H^+\bar{t}P_Rb\,,\nonumber\\[1ex]
  -\left(h_b+\delta h_b\right)
                \frac{\sin\beta}{\sqrt{2}}\,A\bar{b}\gamma_5 b&=&
  -\frac{\mb\tb}{\sqrt{2}\left(1+\Dmb\right)v}\,A\bar{b}\gamma_5 b\,.
\end{eqnarray}
Therefore, in this case, the result of the resummation is to effectively
modify the tree-level Yukawa coupling by the universal $1/(1+\Dmb)$ factor.

\begin{figure}[ht]
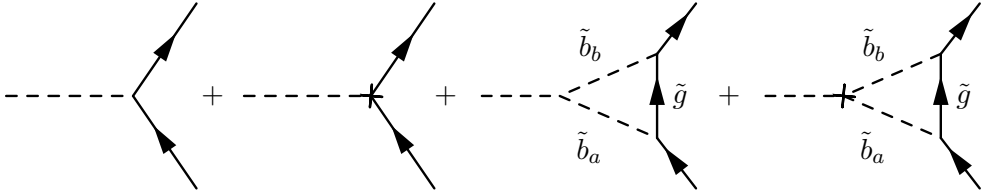

  \begin{center}
\figyukawa
  \end{center}
  \caption{\label{fig:yukawa} Vertex diagrams contributing to the
    renormalization of the Higgs-fermion Yukawa interaction. From left to
    right, the renormalized Yukawa coupling, the Yukawa counterterm, the
    one-loop contribution and the higher-order diagram containing the
    insertion of the $\delta\tilde{h}_b$ counterterm. The solid and dashed
    lines inside the loops denote gluino and sbottom propagators respectively.
    The cross in the fourth diagram denotes the $\delta\tilde{h}_b$
    counterterm.}
\end{figure}

The case of the CP-even neutral Higgs bosons is a little bit more
involved. Depending on the relation between $\beta$ and the
  mixing angle, $\alpha$, the one-loop correction to the vertex
  diagrams can be importantly enhanced.
%In the limit of a heavy \mH, in which the mixing angle, $\alpha$, tends to
%$\beta-\pi/2$, an indirect \tb-enhancement may appear in the vertex diagrams
%through their dependence on $\alpha$. 
The full set of potentially relevant graphs is shown in
fig.~\ref{fig:yukawa}.

One obtains, for the $H\bar{b}b$ renormalized amplitude
\begin{eqnarray}
  \label{eq:effH0Yuk}
-i\left(h_b+\delta h_b\right)\frac{\cos\alpha}{\sqrt{2}}\,H\bar{b}b
  -i\left(\tilde{h}_b+\delta\tilde{h}_b\right)
          \frac{\sin\alpha}{\sqrt{2}}\frac{\Dmb}{\tb}
  \,H\bar{b}b&=&\nonumber\\[1ex]
-i\,\frac{\mb\cos\alpha}{\sqrt{2}\left(1+\Dmb\right)v_1}
  \left(1+\Dmb\frac{\tan\alpha}{\tb}\right)\,H\bar{b}b\,.&&
\end{eqnarray}
Again, the resummation amounts to the inclusion of the universal $1/(1+\Dmb)$
factor. However, there is an additional \Dmb\ term inside the parenthesis,
which constitutes the non \tb-suppressed contribution coming from the SUSY-QCD
vertex diagrams. Similarly, for the $h\bar{b}b$ one has
\begin{eqnarray}
  \label{eq:effh0Yuk}
 i\left(h_b+\delta h_b\right)\frac{\sin\alpha}{\sqrt{2}}\,h\bar{b}b
  -i\left(\tilde{h}_b+\delta\tilde{h}_b\right)
          \frac{\cos\alpha}{\sqrt{2}}\frac{\Dmb}{\tb}
  \,h\bar{b}b&=&\nonumber\\[1ex]
 i\,\frac{\mb\sin\alpha}{\sqrt{2}\left(1+\Dmb\right)v_1}
  \left(1-\frac{\Dmb}{\tan\alpha\tb}\right)\,h\bar{b}b\,.&&
\end{eqnarray}
It can easily be checked that for large \mH\ values, the limit that
corresponds to the effective decoupling of one of the Higgs doublets, one
recovers the SM $h\bar{b}b$ coupling
\[
-i\,\frac{\mb}{\sqrt{2}v}\,h\bar{b}b\,,
\]
whereas the $H\bar{b}b$ coupling, $H$ being heavy, still ``feels'' the
decoupled sector
\[-i\,\frac{\mb\tb}{\sqrt{2}\left(1+\Dmb\right)v}\,H\bar{b}b\,.
\]

\begin{figure}[ht]
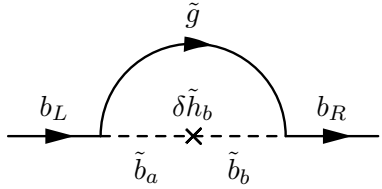

  \begin{center}
    \figextra
  \end{center}
  \caption{\label{fig:extra} 
    The fourth self-energy diagram in fig.~\ref{fig:self}, shown in greater
    detail. A gluino propagator is denoted by the solid line inside the loop.
    The dashed lines denote sbottom propagators and the cross the insertion of
    a $\delta\tilde{h}_b$ counterterm.}
\end{figure}

\setcounter{secnumdepth}{1}
\subsection{Two-point--three-point diagram identity}
\setcounter{secnumdepth}{2}

Let us evaluate the amplitude associated to the three-point Feynman diagram of
fig.~\ref{fig:extra}. Neglecting the external momentum, it can be written
\begin{equation}
  \label{eq:extra}
  -\left(8\pi\aS\right)\,C_F\,\frac{\delta\tilde{h}_b v_1}{\sqrt{2}}
                        \,\mg\,\mu\tb
  \int\frac{d^nk}{(2\pi)^n}\,
  \frac{\left(Z_{i1}Z_{j2}^*+Z_{i2}Z_{j1}^*\right)\,Z_{j2}Z_{i1}^*}
       {(k^2-\mg)(k^2-\msb[2]{i})(k^2-\msb[2]{j})}\,\bar{b}P_Lb\,,
\end{equation}
where $C_F=4/3$ is a colour factor and the two-dimensional rotation matrices
$Z$ transform the weak eigenstate sbottom basis into the mass eigenstate
basis. Expressed in terms of the mixing angle $\theta_{\tilde{b}}$, the
components of $Z$ read: $Z_{11}=Z_{22}=\csb$, $Z_{12}=-Z_{21}=\ssb$. The term
between parentheses in the numerator of~(\ref{eq:extra}) and the combination
$\delta\tilde{h}_b v_2=\delta\tilde{h}_b v_1\tb$ come from the counterterm to
the \tb-dominant interaction $H_2^0\tilde{b}_R^*\tilde{b}_L$, after the Higgs
field develops its vacuum expectation value $v_2$.

Splitting the implicit $i,j$ sum into the $i=j$ part and the rest of the terms
we obtain
\begin{eqnarray}
  \label{eq:split}
  &&\frac{\kappa}{2}\,\sin^2\!2\theta_{\tilde{b}}\int\frac{d^nk}{(2\pi)^n}\,
  \frac{1}{k^2-\mg[2]}
  \left(\frac{1}{(k^2-\msb[2]{1})^2}+\frac{1}{(k^2-\msb[2]{2})^2}\right)
       \,\bar{b}P_Lb\nonumber\\
  &+&\kappa\cos^2\!2\theta_{\tilde{b}}\int\frac{d^nk}{(2\pi)^n}\,
  \frac{1}{(k^2-\mg[2])(k^2-\msb[2]{1})(k^2-\msb[2]{2})}\,\bar{b}P_Lb\,,
\end{eqnarray}
the constant $\kappa$ being a short-hand for the constant prefactor of the
integral in~(\ref{eq:extra}).

The second term in~(\ref{eq:split}) is of the same form as \Dmb.
Adding and removing this term times $\tan^2\!2\theta_{\tilde{b}}$
  and rearranging terms one arrives at
\begin{eqnarray}
  \label{eq:compact}
  &&\kappa\int\frac{d^nk}{(2\pi)^n}\,
  \frac{1}{(k^2-\mg[2])(k^2-\msb[2]{1})(k^2-\msb[2]{2})}\,\bar{b}P_Lb
                \nonumber\\
  &+&\frac{\kappa}{2}\,\sin^2\!2\theta_{\tilde{b}}\int\frac{d^nk}{(2\pi)^n}
  \frac{1}{k^2-\mg[2]}\,
       \frac{(\msb[2]{2}-\msb[2]{1})^2}
                {(k^2-\msb[2]{1})^2(k^2-\msb[2]{2})^2}
       \,\bar{b}P_Lb\,.
\end{eqnarray}
Now one can make use of the tree-level relation
\begin{eqnarray}
\sin\!2\theta_{\tilde{b}} &=&
  \frac{2\mb\left(A_b-\mu\tb\right)}{\msb[2]{1}-\msb[2]{2}}
\label{thbt}
\end{eqnarray}
to write ($A_b$ is dropped since it is not \tb-enhanced)
\begin{eqnarray}
  \label{eq:final}
  &&\kappa\int\frac{d^nk}{(2\pi)^n}\,
  \frac{1}{(k^2-\mg[2])(k^2-\msb[2]{1})(k^2-\msb[2]{2})}\,\bar{b}P_Lb
                \nonumber\\
  &+&2\kappa\,\mb[2]\left(\mu\tb\right)^2\int\frac{d^nk}{(2\pi)^n}
       \frac{1}{(k^2-\mg[2])(k^2-\msb[2]{1})^2(k^2-\msb[2]{2})^2}
       \,\bar{b}P_Lb\,.
\end{eqnarray}

The second integral in (\ref{eq:final}) has two extra propagators and thus in
the limit of heavy SUSY masses it is of ${\cal O}(1/\MSy[6])$, whereas the
first one is of ${\cal O}(1/\MSy[2])$. One can conclude that the two- and
three-point loop diagrams in fig.~\ref{fig:self} are just related by
$\delta\tilde{h}_b/\tilde{h}_b$, apart from contributions that are suppressed
by powers of either \tb\ or $\mb[2]\tb[2]/\MSy[2]$. The amplitude for the
diagram in fig.~\ref{fig:extra} reduces to
\begin{equation}
  \label{eq:limit}
  i\,\delta\tilde{h}_b v_1\,\Dmb
  \left[1+\frac{\mu^2}{\MSy[2]}\,\times
         \,{\cal O}\left(\frac{\mb[2]\tb[2]}{\MSy[2]}\right)
         \,+\,{\cal O}\left(\frac{1}{\tb}\right)\right]\,\bar{b}P_Lb\,.
\end{equation}
 
\newpage

\section{Large logarithms in decay rates}
\label{sec:uli:app}
Our effective lagrangian ${\cal L}$ in eq.~\eq{eq:uli:lag2} contains the large
logarithms associated with the running of the Yukawa couplings to all orders
in perturbation theory. In general this procedure does not sum all the large
logarithms that appear in a specific cross section or decay rate.  In this
appendix we show that for $\Gamma(\tbH)$ and $\Gamma(\Htb)$ such additional,
process-specific logarithms do not occur except in highly power-suppressed,
numerically negligible terms.

Let us first consider the decay $\Htb$: the optical theorem relates the decay
rate to the imaginary part of the $H^+$ self-energy:
\begin{eqnarray}
\Gamma(\Htb) &=& \frac{1}{\mH} \lt. \imag i \int d^4x \, e^{-i q\cdot x} 
        \bra{H^+} {\cal T} J^\dagger(x) J(0) \ket{H^+} 
        \right|_{q^2=\mH[2]}. \label{b:eq:ot}
\end{eqnarray}
Here 
\begin{eqnarray}
J(x) &=& \frac{g}{\sqrt{2} \mw} \frac{\ov{m}_b(Q)}{1+\Dmb} 
         \, V_{tb} \, \tb\, H^+ \,    \ov{t}_L       b_R(x,Q) \no 
\end{eqnarray}
is the scalar current stemming from the Yukawa interaction in
eq.~\eq{eq:uli:lag2}. All currents and couplings in this appendix are
considered to be renormalized using a mass-independent renormalization scheme
such us the $\ov{\rm MS}$ scheme \cite{bbdm}.  For the moment we also assume
this for the quark masses and discuss the use of the pole mass definition,
which is commonly used for the top mass, later.  The decay rate involves
highly separated mass scales $\mb \ll \mH, \mt$. First we assume that $\mH$
and $\mt$ are of similar size so that $\log \mH/\mt$ is not dangerously large.
We return to the case $\mb \ll \mt \ll \mH$ later. To prepare the resummation
of the large logarithm $\log \mb/\mH$, we first perform an operator product
expansion of the bilocal forward scattering operator in eq.~\eq{b:eq:ot}:
\begin{eqnarray}
\!\! i \! \int \!\! d^4x \, e^{-i q\cdot x} 
        \bra{H^+} {\cal T} J^\dagger(x) J(0) \ket{H^+} &=&  
        \sum_n C_n \lt( q^2, \mt, Q \right) 
        \bra{H^+} {\cal O}_n \ket{H^+} \lt( \mb, Q \right) .
        \label{b:eq:ope}
\end{eqnarray}
Here all dependence on the heavy mass scales $\mt$ and $q^2=\mH[2]$ is
contained in the Wilson coefficient $C_n$, while the dependence on the light
scale $\mb$ resides in the matrix element of the local operator ${\cal O}_n$.
Both depend on the renormalization scale $Q$ at which the OPE is carried out
(so that $Q$ is sometimes called \textit{factorization scale}). The OPE
provides an expansion of $\Gamma(\Htb)$ in terms of $(\mb/\mH)^2$. Increasing
powers of $\mb/\mH$ correspond to increasing twists of the local operator
${\cal O}_n$. Here the twist is defined as the dimension of the operator
${\cal O}_n$ minus the number of derivatives acting on the Higgs fields in
${\cal O}_n$.

The OPE in eq.~\eq{b:eq:ope} is depicted in fig.~\ref{b:fig:ope} where
also the leading twist operator ${\cal O}_1=\ov{m}_b^2 (Q) H^+ H^-$ is
shown.  At leading twist the OPE, depicted in fig.~\ref{b:fig:ope}, is
trivial: the matrix element $\bra{H^+} {\cal O}_1 \ket{H^+}$ simply
equals $\ov{m}_b^2(Q)$ and the Wilson coefficient $C_1$ can be read off
from eq.~\eq{eq:hMSSMimp}. In the leading order (LO) of QCD it reads
\begin{figure}[tb]
  \begin{center}
    \figulic
  \end{center}
\caption{The OPE in \eq{b:eq:ope} to leading order in $\mb/\mH$ and
  \aS.  The self-energy diagram on the left represents the left-hand side of
  eq.~\eq{b:eq:ope}.  The right diagram depicts $\bra{H^+} {\cal O}_1
  \ket{H^+}$.  }\label{b:fig:ope}
\end{figure}
\begin{eqnarray}
\imag C_1 &=&
\frac{g^2N_c}{32\pi\mw[2]}\,\mH[2]
  \,\left(1-\rt\right)^2
  \frac{1}{\left(1+\Dmb\right)^2}\,\tb[2]. \label{b:eq:wclo} 
\end{eqnarray}
The QCD radiative corrections in $\Gamma$ contain powers of the large
logarithm $\aS \log \mb/\mH$. The OPE in eq.~\eq{b:eq:ope} splits this
logarithm into $\aS \log Q/\mH + \aS \log \mb/Q $: the former term resides in
the coefficient function $C_1$ while the latter is contained in the matrix
element $\bra{H^+} {\cal O}_1 \ket{H^+}$. If we choose $Q= {\cal O}(
\mt,\mH)$, then the logarithms in the Wilson coefficient are small and
perturbative, but $\log \mb/Q $ in the matrix element is big and needs to be
resummed to all orders. One could likewise choose $Q\simeq \mb$ and resum the
large logarithm in the Wilson coefficient, but the former way is much easier
here. In order to sum $\log \mb/Q $ we have to solve the renormalization group
(RG) equation for ${\cal O}_1$.  Since the Higgs fields in ${\cal O}_1$ have
no QCD interaction, the solution of the RG equation simply amounts to the use
of the well-known result for the running quark mass $\mb(Q)$ (see
eq.~\eq{eq:uli:mass}) at the scale $Q= {\cal O}(\mt,\mH)$ in ${\cal O}_1$. In
the next-to-leading (NLO) order one has to include the ${\cal O} (\aS)$
corrections to $\Gamma$ in eq.~\eq{eq:hMSSMimp}. First there are no explicit
one-loop corrections to $\bra{H^+} {\cal O}_1 \ket{H^+}$, so that in the NLO
we obtain $\imag C_1 (Q)$ by simply multiplying the result in
eq.~\eq{b:eq:wclo} with the curly bracket in \eq{eq:hMSSMimp}. Secondly in the
NLO we have to use the two-loop formula for $\ov{m}_b (Q)$ in the matrix
element.  Since one is equally entitled to use $Q=\mH$ (as chosen in
\eq{eq:hMSSMimp}) or $Q=\mt$ or any other scale of order $\mt,\mH$, there is a
residual scale uncertainty. This feature is familiar from all other
RG-improved observables.  To the calculated order $\aS$ this uncertainty
cancels, because there is an explicit term $\aS \log Q/\mH$ in the one-loop
correction, so that the scale uncertainty is always of the order of the next
uncalculated term. In our case this is ${\cal O}(\aS[2])$ and numerically
tiny. In conclusion, our OPE analysis shows that at leading order in $\mb/\mH$
all large logarithms in $\Gamma(\Htb)$ can indeed be absorbed into the running
quark mass in our effective lagrangian in eq.~\eq{eq:uli:lag2}. Some
clarifying points are in order:
\begin{itemize}
\item[1)] The summation of large logarithms in the NLO does not require the
  calculation of the two-loop diagrams obtained by dressing the diagram in
  fig.~\ref{b:fig:ope} with an extra gluon, as performed in \cite{cd}. This
  calculation only gives redundant information, already contained in the known
  two-loop formula for the running quark mass.
\item[2)] At the next-to-leading order the result depends on the chosen
  renormalization scheme. Changing the scheme modifies the constant term 17/3
  in eq.~\eq{eq:hMSSMimp}. After inserting the NLO (two-loop) solution
  \eq{eq:uli:mass} for the running mass, this scheme dependence cancels
  between this term and $J^{(f)}$ in eq.~\eq{eq:uli:dj}. In the literature,
  sometimes, the one-loop result for $\Gamma$ is incorrectly combined with the
  one-loop running bottom mass resulting in a scheme-dependent
  expression.\\
  No running top-quark mass is needed for the case $\mt \simeq \mH$, and one
  can adopt the pole mass definition for \mt\ as we did.
\item[3)] The OPE also shows that the correct scale to be used in the running
  $\aS$ in eq.~\eq{eq:hMSSMimp} is the high scale $Q={\cal O} (\mt,\mH)$ and
  not the low scale \mb.
\item[4)] The absorption of the large logarithms into the running mass does
  not work for terms that are suppressed by higher powers $(\mb/\mH)^n$ with
  respect to the leading contribution considered by us. Higher-twist operators
  contain explicit $b$-quark fields.  At twist-8 there are operators of the
  form $\mb[3] H^+ H^- \ov{b} \Delta b $, where $\Delta$ is some Dirac
  structure. Solving the RG equation for these operators yields extra
  evolution factors in addition to the running mass. These effects occur in
  corrections of order $\mb[4]/\mH[4]$ and are certainly only of academic
  interest.
\end{itemize}
Next consider $\Gamma (\Htb)$ for the case $\mb \ll \mt \ll \mH$: in this
limit, another large logarithm, $\log \mt/\mH$, appears. Now we have to
perform the OPE in two steps. In the first step we again match the forward
scattering operator to local operators as in eq.~\eq{b:eq:ope} at a scale
$Q_1={\cal O} (\mH)$, but we treat the top quark as light, so that the
dependence on \mt\ now resides in the matrix element $\bra{H^+} {\cal O}_1
\ket{H^+}$ rather than in the Wilson coefficient. For simplicity we specify
$Q_1=\mH$.  The leading power $\rt[0]$ is again represented by the twist-4
operator $ {\cal O}_1$, yet the corresponding Wilson coefficient lacks the
factor of $ (1-\rt)^2$ compared to eq.~\eq{b:eq:wclo}.  The terms of order
$\rt[1]$ are represented by $ {\cal O}_2 = \ov{m}_t^2 {\cal O}_1$ with
$C_2=-2\, C_1/\mH[2]$.  At twist-8 different operators of the form $\mt[3] H^+
H^- \ov{t} \Delta t $ with non-trivial anomalous dimensions occur as discussed
in point 4 above. In the second step one applies an OPE at the scale $Q_2 =
{\cal O} (\mt)$. At this step the dependence on \mt\ migrates from the matrix
elements into the Wilson coefficients, which at order $\rt[1]$ amounts to a
trivial rescaling of the coefficients and operators by \mt\ or $1/\mt$. To
order $\rt[0]$ and $\rt[1]$ the only effect of the OPE is to replace the top
mass in the expression for $\Gamma (\Htb)$ in eq.~\eq{eq:hMSSMimp} by a
running top mass $\ov{m}_t(\mH)$, and to omit the explicit term proportional
to $r_t \log r_t$ in the ${\cal O}(\aS)$ correction. Since we have adopted the
on-shell definition for the top mass, one must either use a running mass
definition based on the pole mass (i.e.\ with $\ov{m}_t(\mt)=m_t^{pole}$) or
transform the result in eq.~\eq{eq:hMSSMimp} to the $\ov{\rm MS}$ scheme with
the appropriate change in the ${\cal O}(\aS)$ correction. It is a nice check
to expand the running mass to first order in \aS:
\begin{eqnarray}
\ov{r}_t \lt( \mH \right) &=& \frac{\ov{m}_t^2(\mH)}{\mH[2]} \; = \;
                \rt  \,  
        \lt( 1 + 2\, \frac{\aS}{\pi} \log \rt   \right) + 
        {\cal O} \lt( \aS[2] \right)
\qquad \qquad \mbox{with } \rt = \ov{r}_t \lt( \mt \right) \no
\end{eqnarray}
and to verify that the overall factor $(1-\ov{r}_t ( \mH ) )^2$ indeed
reproduces the $\rt \log \rt$ term in eq.~\eq{eq:hMSSMimp}. The terms of order
$\rt[2] \log \rt$ are not correctly reproduced by the running top mass as
anticipated by the occurrence of non-trivial twist-8 operators.  The important
result of our consideration of the case $\mb \ll \mt \ll \mH$ is the absence
of terms of the form $\rt[0] \log \rt$ to all orders in \aS. In this case the
additional large logarithm $\log r_t$ is always suppressed by powers of $r_t$
and therefore these terms are negligible for $\rt \ll 1$ and need not be
resummed.

For the decay $\tbH$ the above discussion can be repeated with the appropriate
changes in the OPE: the leading-twist operator is now ${\cal O}_1
(Q)=\ov{m}_b^2 (Q) \, \ov{t} t (Q)$ and the external state in
eq.~\eq{b:eq:ope} is a top quark instead of a charged Higgs boson.  We have
$\mb \ll \mt,\mH$ and the factorization scale $Q$ is again of order $\mt,\mH$.
While ${\cal O}_1$ now involves strongly interacting fields, its matrix
element $\bra{t}{\cal O}_1\ket{t}(Q) $ still does not contain large logarithms
$\log \mb/Q$ other than those contained in the running mass $\ov{m}_b (Q)$.
Hence the proof above for $\Gamma (\Htb)$ applies likewise for $\Gamma
(\tbH)$.

After exchanging $V_{tb} \ov{t}_L b_R$ for $V_{cb} \ov{c}_L b_R$ in
eq.~\eq{eq:uli:lag2}, we can likewise apply our effective lagrangian ${\cal
  L}$ to semileptonic $B$-meson decays corresponding to $b \rightarrow c\,
\ell \, \ov{\nu}_{\ell}$ by using the appropriate scale $Q\simeq \mb$ in
${\cal L}$.  The QCD radiative corrections involve no large logarithm, because
the gluons couple only to the $b$ and $c$ quarks. Hence the effective
four-fermion operator $\ov{c}_L b_R \, \ov{\ell}_R \nu_{L}$ obtained after
integrating out the heavy $H^+$ renormalizes in the same way as the quark
current $ \ov{c}_L b_R$ in ${\cal L}$. The corresponding loop integrals do not
depend on \mH\ at all and this feature is correctly reproduced by using
$\ov{m}_b (\mb)$ in ${\cal L}$. The situation is different in physical
processes in which the charged Higgs connects two quark lines, as for example
in the loop-induced decay $b\rightarrow s \, \gamma$. Here effective
four-quark operators, which involve a non-trivial renormalization group
evolution, occur. The large-\tb\ supersymmetric QCD corrections associated
with $\Dmb$ and the $H^+\ov{t}_L b_R $ Yukawa coupling, however, are still
correctly reproduced by applying ${\cal L}$ to $b\rightarrow s \, \gamma$ or
other loop-induced rare $b$-decays. Yet it must be clear that these
corrections are part of the mixed electroweak-QCD two-loop contributions and
that there are already supersymmetric electroweak contributions at the
one-loop level, which are process-specific and of course not contained in
${\cal L}$.

%%% Local Variables: 
%%% mode: latex
%%% TeX-master: "eff"
%%% End: 

%%%%%%%%%%%%%%%%%%%%%%%%%%%%%%
\end{document}